\documentclass[12pt]{article}
\pdfoutput=1
\usepackage{graphicx}
\usepackage{amsfonts}

\usepackage[T1]{fontenc}
\usepackage[latin9]{inputenc}
\usepackage{geometry}
\usepackage{amstext}
\usepackage{amssymb}

\usepackage{setspace}
\def\hbar{\hspace{0pt}\raisebox{1pt}{$-$} \hspace{-7pt} h}

\def\5{\overline 5}

\newcommand{\be}{\begin{equation}}
\newcommand{\ee}{\end{equation}}
\newcommand{\bea}{\begin{eqnarray}}
\newcommand{\eea}{\end{eqnarray}}

\usepackage{slashed}

\begin{document}
\title{
\vspace{-1cm}
\bf A functional RG equation\\ for the $c$--function \vspace{1cm}}
\vspace{3cm}
\author{Alessandro Codello$^{a}$, Giulio D'Odorico$^{a}$ and Carlo Pagani$^{a,b}$\\
$^{a}$\emph{SISSA, via Bonomea 265, 34136 Trieste, Italy}\\
$^{b}$\emph{INFN, Sezione di Trieste}
\vspace{1cm}}
%
%

\date{}
\maketitle

\begin{abstract}

After showing how to prove the integrated $c$--theorem within the functional RG framework based on the effective average action,
we derive an exact RG flow equation for Zamolodchikov's  $c$--function in two dimensions by relating it to the flow of the effective average action.
In order to obtain a non--trivial flow for the $c$--function,
we will need to understand the general form of the effective average
action away from criticality, where nonlocal invariants, with beta functions as coefficients, must be included in the ansatz to be consistent.
Then we apply our construction to several examples: exact results, local potential approximation and loop expansion.
In each case we construct the relative approximate $c$--function and find it to be consistent with Zamolodchikov's 
$c$--theorem.
Finally, we present a relation between the $c$--function and the (matter induced) beta function of Newton's constant,
allowing us to use heat kernel techniques to compute the RG running of the $c$--function. 

\end{abstract}

%
\vskip1.5em


\tableofcontents{}

\newpage

\section{Introduction}

The renormalization group (RG) underlies most of our modern understanding
of quantum and statistical field theories \cite{Wilson:1974mb}.
There are different ways to implement the RG procedure. Whereas standard
sliding scale arguments (Gell-Mann-Low) are particularly suitable
for weakly coupled computations, it is only with Wilson's ideas that
non-perturbative insights have been possible.

The arena where the RG acts is theory space. This space is parametrized by all couplings
corresponding to terms which are consistent
with the symmetries of the system we want to study. The beta functions for the couplings define a vector
field in theory space, and the RG flow can be seen in geometrical
terms as a certain trajectory in this space.

From this point of view the infrared physics depends upon the differential
equation governing the flow as well as on the boundary conditions. 
If the initial point sits at
a finite scale (for example, it is a bare action depending on some
UV cutoff $\Lambda$), one is considering an effective field
theory, whose range of validity is limited by the cutoff scale.
However, if we want a theory to be called fundamental,
we would like to be able to push the initial scale to arbitrarily
high values, eventually to infinity. The only known way to perform
this limit is to hit a UV fixed point.

Fixed point theories do not depend on any intrinsic scale since they are scale invariant.
As a consequence they can be used to model systems at criticality. 
These theories are characterized by dimensionless couplings and  physical quantities 
exhibit scaling relations which can be observed in experiments. 
These relations arise in very different systems
sharing the same dimensionality, symmetry and field content. This is what
is usually referred to as the concept of universality, the independence
of the critical properties of a system from its microscopic details.
The RG offers a simple and intuitive explanation of universality:
the critical properties of a system are determined by the fixed
point, microscopic actions defined at different scales
that flow to the same fixed point, or equivalently that belong to
the same basin of attraction of a fixed point, will describe the same
criticality.

We see that in this light the problem of understanding the critical
properties realized in nature boils down to the classification of all the different fixed points.
In two dimensions we know that every unitary scale invariant theory is also
conformal invariant, so the problem further reduces to the classification
of all possible conformal field theories (CFT).
This can be done via algebraic methods, exploiting the properties of the associated Virasoro
algebra \cite{Belavin:1984vu,mussardo}.


A fixed point theory can then be deformed by adding weakly
coupled operators that trigger a nontrivial flow out of the fixed
point. By considering the linearization of this flow we can obtain
all the remaining CFT data (like scaling dimensions and other critical
exponents) that characterize the physical system and the way in which
it responds to deformations. This is also the main idea of conformal
perturbation theory.

So far our discussion has been limited to the neighborhood of a fixed point.
The next natural step is to try to gain more information on the global
properties of theory space. 
Such information is provided by Zamolodchikov's $c$--theorem \cite{Zamolodchikov:1986gt}, which states that in every
unitary Poincar\'e invariant theory there exists a function of the coupling constants,
the $c$--function, that decreases from CFT$_{UV}$ to CFT$_{IR}$, and
that is stationary at the endpoints of the flow, where its value equals
the central charge of the corresponding CFT.
Note that the difference between the
two central charges is an intrinsic quantity (intrinsic meaning independent of spurious contributions
like scheme dependence of the renormalization procedure), so the content of
the theorem is highly nontrivial.

In this case a complete RG analysis  requires the ability to follow the flow arbitrarily far
away from a fixed point. Unless the two fixed points are sufficiently
close to each other we cannot rely on perturbative schemes. The non-perturbative framework
we will use to address these issues is the functional renormalization group (fRG)
based on the effective average action (EAA) \cite{Berges:2000ew}.
The EAA is a functional whose scale dependence is given by an exact flow equation \cite{Wetterich:1992yh}
which, being exact, allows to explore non-perturbative aspects. 
A first application of exact RG equations to the $c$--function has been explored in \cite{Generowicz:1997he,Haagensen:1993by}.
%


The main purpose of this paper is to move the first steps necessary
in order to give a bridge between these two general results: the $c$--theorem
and the computation of universal quantities related to the integrated
flow between fixed points (that is, to global properties of theory
space), and the fRG formalism based on the exact flow for the EAA.
Our approach will be mainly a constructive one. We will give a general
recipe to construct a $c$--function compatible with Zamolodchikov's
theorem within the fRG framework.
After identifying a natural candidate for a scale dependent
$c$--function, $c_{k}$,  we will be able to
write an exact non-perturbative flow equation for it.
%
%
Of course, there are only few cases in which the exactness of the flow equation
can be used and one usually needs to resort to approximations.
However, we will see that already for a simple truncation as the local potential approximation
the flow equation gives results compatible with the $c$--theorem.

Our viewpoint will be based upon a curved space construction.
The reason for this is twofold. First, this avoids having to resort
to algebraic techniques or OPE analysis: the central charge, for instance,
becomes the coefficient of the conformal anomaly, which in curved
space becomes an operator anomaly in the one point function.
Second, this is more suitable
for functional techniques, as the derivation of the trace anomaly
matching condition will show, and more useful to write a general effective
action. Indeed, this construction will require an investigation of
what is the general form of the EAA away from fixed points,
since the usual expansion in local operators is incapable of giving
a nonzero running for $c_k$.
Working in curved space the natural candidate for $c_{k}$ is the coefficient of
the Polyakov action. We will take this as our definition for the $c$--function and leave
for further study the mapping between our approach and the one based on local
RG with spacetime dependent couplings \cite{Osborn:1991gm}. 
%


The paper will be organized as follows.
In section 2 we will construct a Weyl--invariant functional measure and discuss the form of a CFT on curved background. 
This will lead us to a re--derivation of the trace anomaly matching condition,
from which the {}``integrated'' $c$--theorem follows from known results
\cite{Komargodski:2011xv}.
We will then move on to discuss the scale dependent
$c$--function, and obtain our flow equation for it, in section 3. This
construction uses the EAA as the main tool, so
in section 4 we will investigate
its general form.
In section 5 we discuss various applications of our formalism 
while in section 6 we put forward a simple relation between the beta function
of Newton's constant and the running $c$--function.
%
Section 7 is devoted to the conclusions.

\section{The integrated $c$--theorem}

We start by reviewing the integrated $c$--theorem expressing the
change of the central charge $\Delta c=c_{UV}-c_{IR}$ along a RG
trajectory connecting two fixed point theories, or equivalently two CFTs.
We will work in curved space where the central charge, or equivalently
the conformal anomaly, can be seen as the coefficient of the Polyakov
term in the effective action. When we specify the background metric
to be of the specific form $g_{\mu\nu}=e^{2\tau}\delta_{\mu\nu}$,
with $\tau$ the {}``dilaton'', $\Delta c$ becomes the coefficient
of the operator $\int \tau\Delta\tau$ and can be easily
extracted. But before we need to briefly discuss functional measures
in curved space, Weyl--invariant quantization and the form of the effective action for a CFT on a curved background.


\subsection{Weyl--invariant quantization and functional measures}

The standard diffeomorphism invariant path integral
measure in curved space \cite{Mottola:1995sj}, denoted here ${\cal D}_{g}^{I}$, is Weyl--anomalous:
under a Weyl transformation of the background metric $g_{\mu\nu}\rightarrow e^{2\tau}g_{\mu\nu}$
and of the fields $\phi\to e^{w\tau}\phi$,
where $w$ is the conformal weight of the field\footnote{For a scalar field $w_{\phi}=-\left(\frac{d}{2}-1+\frac{\eta_\phi}{2}\right)$, while for a fermion field $w_{\psi}=-\left(\frac{d}{2}-\frac{1}{2}+\frac{\eta_\psi}{2}\right)$. The conformal weight of the metric is $w_g=2$ in every dimension.},
one encounters the conformal anomaly:
\begin{equation}
{\cal D}_{e^{2\tau}g}^{I}\left(e^{w\tau}\phi\right)={\cal D}_{g}^{I}\phi\, e^{-c\,\Gamma_{WZ}[\tau,g]}\,,\label{WZ}
\end{equation}
where $c$ is the central charge of the CFT, which we want to use as UV action in the path integral,
and $\Gamma_{WZ}[\tau,g]$ is the Wess--Zumino action:  										           
\begin{equation}
\Gamma_{WZ}[\tau,g]=-\frac{1}{24\pi}\int d^{2}x\sqrt{g}\left[\tau\Delta\tau+\tau R\right]\,,\label{1.21}
\end{equation}
where $\Delta\equiv -\nabla_\mu \nabla^\mu$ is the Laplacian.

The Wess--Zumino action can be integrated to give the related Polyakov action, 
\be
S_{P}[g]=-\frac{1}{96\pi}\int d^{2}x\sqrt{g}R\frac{1}{\Delta}R\,,
\ee
which, upon Weyl variation, gives back (\ref{WZ}):
\begin{equation}
S_{P}[e^{2\tau}g]-S_{P}[g]=\Gamma_{WZ}[\tau,g]\,.\label{3}
\end{equation}
The Polyakov action generates the following
quantum energy--momentum tensor,
\begin{eqnarray*}
\left\langle T^{\mu\nu}\right\rangle  & = & \frac{c}{48\pi}\left[-2\nabla^{\mu}\nabla^{\nu}\frac{1}{\Delta}R-\left(\nabla^{\mu}\frac{1}{\Delta}R\right)\left(\nabla^{\nu}\frac{1}{\Delta}R\right)+\right.\\
 &  & \qquad\qquad\qquad\qquad\qquad\qquad\qquad\,\left.-2g^{\mu\nu}R+\frac{1}{2}g^{\mu\nu}\left(\nabla^{\alpha}\frac{1}{\Delta}R\right)\left(\nabla_{\alpha}\frac{1}{\Delta}R\right)\right]\,,
\end{eqnarray*}
%
which is anomalous:											
\begin{equation}
\left\langle T_{\,\mu}^{\mu}\right\rangle =-\frac{c}{24\pi}R\,.
\label{10}
\end{equation}
This is the conformal anomaly, in the two dimensional case.
In curved space, where it can be written in terms of curvature invariants, the conformal anomaly manifests itself already in the one--point function (\ref{10}),
while in flat space it is seen only starting from the two--point function.
For example, in flat space the two point function of the energy--momentum
tensor obtained from the Polyakov action, when written in complex coordinates,
reproduces the standard CFT result \cite{Belavin:1984vu}:
\be
\left\langle T_{zz}T_{ww}\right\rangle =\frac{1}{(2\pi)^{2}}\frac{c/2}{(z-w)^{4}}\,.
\ee
This relation shows the equivalence between the central charge and anomaly coefficient.

We can use the Polyakov action to define, formally, a new measure in the following way:
\begin{equation}
{\cal D}_{g}^{II}\phi\equiv{\cal D}_{g}^{I}\phi\, e^{cS_{P}[g]}\,.\label{1}
\end{equation}
Now using (\ref{WZ}) and (\ref{3}) one can show that indeed (\ref{1}) is Weyl--invariant:
\begin{eqnarray}
{\cal D}_{e^{2\tau}g}^{II}\left(e^{w\tau}\phi\right) & = & {\cal D}_{e^{2\tau}g}^{I}\left(e^{w\tau}\phi\right)\, e^{cS_{P}[e^{2\tau}g]}\nonumber \\
 & = & {\cal D}_{g}^{I}\phi\, e^{-c\Gamma_{WZ}[\tau,g]}e^{cS_{P}[g]+c\Gamma_{WZ}[\tau,g]}\nonumber \\
 & = & {\cal D}_{g}^{II}\phi\,.\label{4}
\end{eqnarray}
With these definitions, we now look at the effective action.
First we define the standard Weyl non--invariant effective action\footnote{We define $\int_{1PI}\equiv\int e^{\int \sqrt{g} \,\Gamma^{(1,0)}[\varphi,g]\chi}$ where $\Gamma^{(1,0)}[\varphi,g]\equiv \frac{\delta \Gamma [\varphi,g]}{\delta \varphi}$.}:					
\begin{equation}
e^{-\Gamma_{I}[\varphi,g]}=\int_{1PI}{\cal D}_{g}^{I}\chi\, e^{-S[\varphi+\chi,g]}\,.\label{5}
\end{equation}
If the bare or UV  action is conformally invariant $S[e^{w\tau}\phi,e^{2\tau}g]=S[\phi,g]$,
this is not so for the standard effective action, which instead satisfies the Wess--Zumino relation:
\begin{equation}
\Gamma_{I}[e^{w\tau}\varphi,e^{2\tau}g]-\Gamma_{I}[\varphi,g]=c\Gamma_{WZ}[\tau,g]\,.\label{WZR}
\end{equation}
Using instead the Weyl--invariant measure defined in (\ref{1}) to define the effective action,  															
\begin{equation}
e^{-\Gamma_{II}[\varphi,g]}=\int_{1PI}{\cal D}_{g}^{II}\chi\, e^{-S[\varphi+\chi,g]}\,,   \label{7}     
\end{equation}
gives rise to a Weyl--invariant effective action:
\begin{equation}
\Gamma_{II}[e^{w\tau}\varphi,e^{2\tau}g]=\Gamma_{II}[\varphi,g]\,.
\label{8}
\end{equation}
Equation (\ref{8}) is valid only when $\Gamma_{II}[\varphi,g] =S[\varphi,g]$, 
%
but still is important from the RG point of view: it is possible to obtain a Weyl--invariant effective
action only if there are no perturbations to the UV action and thus no induced RG flow.
Thus the (bare) UV action and the (effective) IR action are the same in this case.
Said in other words, the path integration amounts to the substitution of the quantum field with the average field.
A purely Gaussian theory provides an example where one can check explicitly the validity of equation (\ref{8}).

Similar reasoning has been made in \cite{Codello:2012sn} with the exception that in that work a St\"uckelberg trick was used to maintain Weyl--invariance for any UV action.

\subsection{CFT action on curved background}

We have seen how to define, at least formally, a Weyl--invariant effective action starting from a Weyl invariant UV action via the functional measure (\ref{1}), which is to be understood as the measure we will use from now on.
Nevertheless on a curved background the effective action of a CFT is not Weyl--invariant
since every CFT with $c \neq 0$ is anomalous, and thus its action must contain a Polyakov term.
Still, in absence of relevant perturbations, quantization will just give the IR effective action equal to the UV action. 

These considerations lead to the following {}``split'' form for the effective action of a general CFT in presence of  a background metric:
%
%
\begin{equation}
\Gamma[\phi,g]=S_{CFT}[\phi,g]+cS_{P}[g]\,.\label{9}
\end{equation}
Here $S_{CFT}[\phi,g]$ is the curved space generalization of the flat space CFT action $S_{CFT}[\phi]\equiv S_{CFT}[\phi,\delta]$, defined by its Taylor series expansion in terms of correlation functions of $\phi$, these being, in principle, exactly known.
Very few CFT actions can be written in local form, these are the Gaussian, the Ising model (in the fermion representation) and the Wess--Zumino--Witten AKM actions \cite{mussardo}.
$S_{P}[g]$ is the Polyakov action and $c$ its central charge.
Other possible Weyl--invariant terms depending on the metric alone are not present in $d=2$, but appear in higher dimensions. 

We now give an explicit example of this construction.
The Gaussian theory has $c=1$ and is the simplest example of a CFT:
\begin{equation}
S_{CFT}^{c=1}[\phi,g]=\frac{1}{2}\int\sqrt{g}\phi\Delta\phi
\,.\label{11}
\end{equation}
%
Using the one--loop trace--log formula starting from the Gaussian UV action $\Gamma_{UV}$ we find:
\be
\Gamma_{IR}[\phi,g] =  \Gamma_{UV}[\phi,g]+\frac{1}{2}\textrm{Tr}\log\Delta-S_{P}[g] =  \Gamma_{UV}[\phi,g]\,,
\ee
where the second term is due to the integration of the fluctuations,
while the Polyakov term with the minus sign comes from the Weyl--invariant measure (\ref{1}).
The two cancel since the $\frac{1}{2}\textrm{Tr}\log\Delta=S_{P}[g]$.
In order to have $\Gamma_{UV} \neq \Gamma_{IR}$ one needs to add a relevant perturbation triggering the RG flow.

%

\subsection{Anomaly matching from the path-integral}


%
%
Starting from $\Gamma_{UV}[\phi,g]=S_{UV}[\phi,g]+c_{UV}S_{P}[g]$
plus relevant operators, we can consider the IR effective action
obtained by integrating out fluctuations:
\begin{eqnarray}
e^{-\Gamma_{IR}[\varphi,g]} & = & \int_{1PI}{\cal D}_{g}\chi\, e^{-S_{UV}[\varphi+\chi,g]-c_{UV}S_{P}[g]+relevant}\nonumber\\
 & = & e^{-c_{UV}S_{P}[g]}\int_{1PI}{\cal D}_{g}\chi\, e^{-S_{UV}[\varphi+\chi,g]+relevant}\,.
\label{anomaly_match}
\end{eqnarray}
Since the metric is non--dynamical we passed the Polyakov term through the path integral.
Here by \emph{relevant} we mean, depending on the case, massive deformations or marginally relevant ones. An example of the first are mass terms like $\frac{m^2}{2}\phi^2$ or $m\bar{\psi}\psi$, while Yang--Mills theory is an example of the second case.  

If we now flow to an IR fixed point,
by virtue of the splitting property (\ref{9}), we must have $\Gamma_{IR}[\phi,g]=S_{IR}[\phi,g]+c_{IR}S_{P}[g]$.
Choosing a dilaton background of the form $g_{\mu\nu}=e^{2\tau}\delta_{\mu\nu}$, we are left with:
\begin{equation}
e^{-S_{IR}[\varphi,e^{2\tau}\delta]}e^{(c_{UV}-c_{IR})\Gamma_{WZ}[\tau,\delta]}=\int_{1PI}{\cal D}_{e^{2\tau}\delta}\chi\, e^{-S_{UV}[\varphi+\chi,e^{2\tau}\delta]+relevant}\,,
\end{equation}
where we used (\ref{3}) on flat space $\Gamma_{WZ}[\tau,\delta]=S_{P}[e^{2\tau}\delta]$.
In order to recover the flat space measure we first shift $\chi\to e^{w\tau}\chi$ and $\varphi\to e^{w\tau}\varphi$ and then use the invariance (\ref{1}):
\begin{equation}
e^{-S_{IR}[e^{w\tau}\varphi,e^{2\tau}\delta]}e^{(c_{UV}-c_{IR})\Gamma_{WZ}[\tau,\delta]} = \int_{1PI}{\cal D}_{\delta}\chi
\,e^{-S_{UV}[e^{w\tau}(\varphi+\chi),e^{2\tau}\delta]+relevant}\,.
\end{equation}
Then we use the conformal invariance properties of the actions, i.e. we substitute $S_{UV}[e^{w\tau}\phi,e^{2\tau}\delta]=S_{UV}[\phi]$ and $S_{IR}[e^{w\tau}\phi,e^{2\tau}\delta]=S_{IR}[\phi]$ since both actions are Weyl--invariant:
\begin{eqnarray}
e^{-S_{IR}[\varphi]}e^{(c_{UV}-c_{IR})\Gamma_{WZ}[\tau,\delta]} & = & \int_{1PI}{\cal D}_{\delta}\chi\, e^{-S_{UV}[\chi+\varphi]+relevant}
\,.
\label{13}
\end{eqnarray}
Note that ${\cal D}_{\delta}\chi\equiv{\cal D}\chi$ is the flat space measure.
The only remaining dependence on $\tau$ is due to the relevant terms, which make the path integral non--trivial.
The last equality 
tells us
that the dilaton effective action (generated by matter loops)
compensates exactly the difference between the anomalies in the UV
and IR.
This is precisely the anomaly matching condition
considered in
\cite{Komargodski:2011xv,Schwimmer:2010za}.
%

\subsection{Proof of the integrated $c$--theorem}

We can now prove the integrated $c$--theorem following \cite{Komargodski:2011xv}. From equation (\ref{13}),
\begin{equation}
e^{-S_{IR}[\varphi]}e^{-\frac{c_{UV}-c_{IR}}{24\pi}\int \tau \Delta \tau}=\int_{1PI}{\cal D}\chi\, e^{-S_{UV}[\varphi+\chi]+relevant}\,,  \label{14}
\end{equation}
we can read off $\Delta c$ from the terms of the dilaton two--point function								
quadratic in momenta.
The \emph{relevant} terms can be expanded in powers of $\tau$:
\be
relevant =  \int d^{2}x\,\tau \,\Theta 
+O(\tau^2)\,,
\label{15}
\ee
where $\Theta \equiv T_{\mu}^{\mu}$ and we omitted all terms of order $\tau^2$ or greater since it is easy to see that they will not contribute to $\int \tau \Delta \tau$. 
%
We are thus interested in the following expectation:
\be
\left.\left\langle e^{\int\tau \,\Theta}
\right\rangle \right|_{\tau^{2}}=
\frac{1}{2}\int d^{2}x\int d^{2}y\,\tau_{x}\tau_{y}\left\langle \Theta_{x}\Theta_{y}\right\rangle
\,.
\ee
%
We now only have to expand $\tau_{y}$ around $\tau_{x}$:
\be
\tau_{y}=\tau_{x}+\left(y-x\right)^{\mu}\partial_{\mu}\tau_{x}+\frac{1}{2}\left(y-x\right)^{\mu}\left(y-x\right)^{\nu}\partial_{\mu}\partial_{\nu}\tau_{x}+...\,;
\label{tauexp}
\ee
%
use translation invariance and compare with the coefficient of $\int \tau\Delta\tau$ to find:
\begin{equation}
\Delta c=3\pi\int d^{2}x\, x^{2}\,\left\langle \Theta_x\Theta_0\right\rangle_{IR} \,,\label{ct}
\end{equation}
which is the integrated version of the $c$--theorem.
From here one simply notices that the integral is positive due to reflection positivity and concludes that $\Delta c \geq 0$ \cite{mussardo,Zamolodchikov:1986gt}. The above equation has also been found with different techniques and expressed as sum rule \cite{Cappelli:1990yc}. 

%

\section{Flow equation for the $c$--function}

The $c$--theorem states \cite{Zamolodchikov:1986gt} that for a two--dimensional unitary quantum field theory, invariant under rotations and
whose energy--momentum tensor is conserved, there exists a function $c$ of the coupling constants which is monotonic along the RG flow and, at a fixed point, is
stationary and equal to the central charge of the corresponding CFT.
Therefore this function $c$ is such that $ \partial_t c<0$
(where the {}``RG time'' is given by the logarithm of the radius $t=\log r$,
so the flow is towards the infrared for $r \to \infty$, hence the minus sign). The differential equation for $c$ can be integrated from
$r=0$ to $r=\infty$ and gives back (\ref{ct}). A natural trial definition for an interpolating $c$--function is given by taking
(\ref{ct}) with the integral which has been cut off at some scale $\mu$ (see for instance \cite{Friedan:2009ik}):
\be
\Delta c\left(\mu\right)\equiv c_{UV}-c\left(\mu\right)=3\pi\int_{0}^{2\pi}d\varphi\int_{0}^{\mu^{-1}}dr\, r^{3}\,\left\langle \Theta(r)\Theta(0)\right\rangle \,.
\ee
We will follow a different approach. Instead of cutting off directly in real space we will cutoff in momentum space.
This will allow us to naturally connect with the framework of the functional Renormalization Group (fRG) and 
to derive an exact RG flow equation for the $c$--function. 
%


\subsection{The fRG flow equation for the $c$--function}

One way to construct the $c$--function is to consider a Wilsonian RG prescription.
A clever way to do the momentum shell integration in a smooth way,
is to introduce a suppressing factor in the path integral
via $\mathcal{D}_g\chi \to \mathcal{D}_g\chi \, e^{-\Delta S_{k}[\varphi,g]}$.
The role of the cutoff action $\Delta S_{k}[\varphi,g]$ is to restrict the integration to modes
above the IR scale $k$.
In this way we obtain a scale dependent effective action $\Gamma_k[\varphi,g]$,
which, using (\ref{9}), can be decomposed as:
\be
\Gamma_k[\varphi,g] = S_k[\varphi,g]+c_k S_P[g] + \textrm{gravitational terms}\,.
\label{splitk} 
\ee
where $S_k[\varphi,g]$ is defined by $S_k[0,g]=0$ and $c_k$ is the scale dependent $c$--function.
By "gravitational terms" we mean the purely geometrical terms depending on the metric alone, like $\int\sqrt{g}$ or $\int\sqrt{g}R$, generated by fluctuations.    
The collection of the $\Gamma_k[\varphi,g]$ for all $k$ constitute the RG trajectory connecting $\Gamma_{UV}[\varphi,g]$ to $\Gamma_{IR}[\varphi,g]$; a cartoon of this shown in figure \ref{fig:  cartoon}.
If we now repeat the steps leading to equation (\ref{14}), but with the cutoff term added, we arrive at:
%
%
\be
e^{-S_{k}[\varphi,e^{2\tau}\delta]}
e^{-\frac{c_{UV}-c_k}{24\pi} \int \tau \Delta \tau}=
\int_{1PI} {\cal D}\chi\, e^{-S_{UV}[\varphi+\chi]+relevant}e^{-\Delta S_{k}[e^{w\tau}\chi,e^{2\tau}\delta]}\, 
\,. \label{defck}
\ee
%
Now a derivative of (\ref{defck}) with respect to the {}``RG time'' $t=\log k$
gives the RG flow of the central charge:
%
\be
\partial_{t}c_k = -24\pi \left\langle \partial_{t}\Delta S_{k}[e^{w\tau}\chi,e^{2\tau}\delta]\right\rangle\Big|_{\int \tau \Delta \tau}\,,
\label{dtc}
\ee
in which the expectation value is calculated within the regularized path integral.
We see that we obtain the flow of the $c$--function if we are able to evaluate the r.h.s. of (\ref{dtc}), after specifying the form of the cutoff action. 
%
%
%
The running of $c_{k}$ is related to the coarse--grained dilaton  two--point function.
To understand how to handle this equation, we need to introduce the effective average action (EAA).
\begin{figure}
\begin{centering}
\includegraphics[scale=0.2]{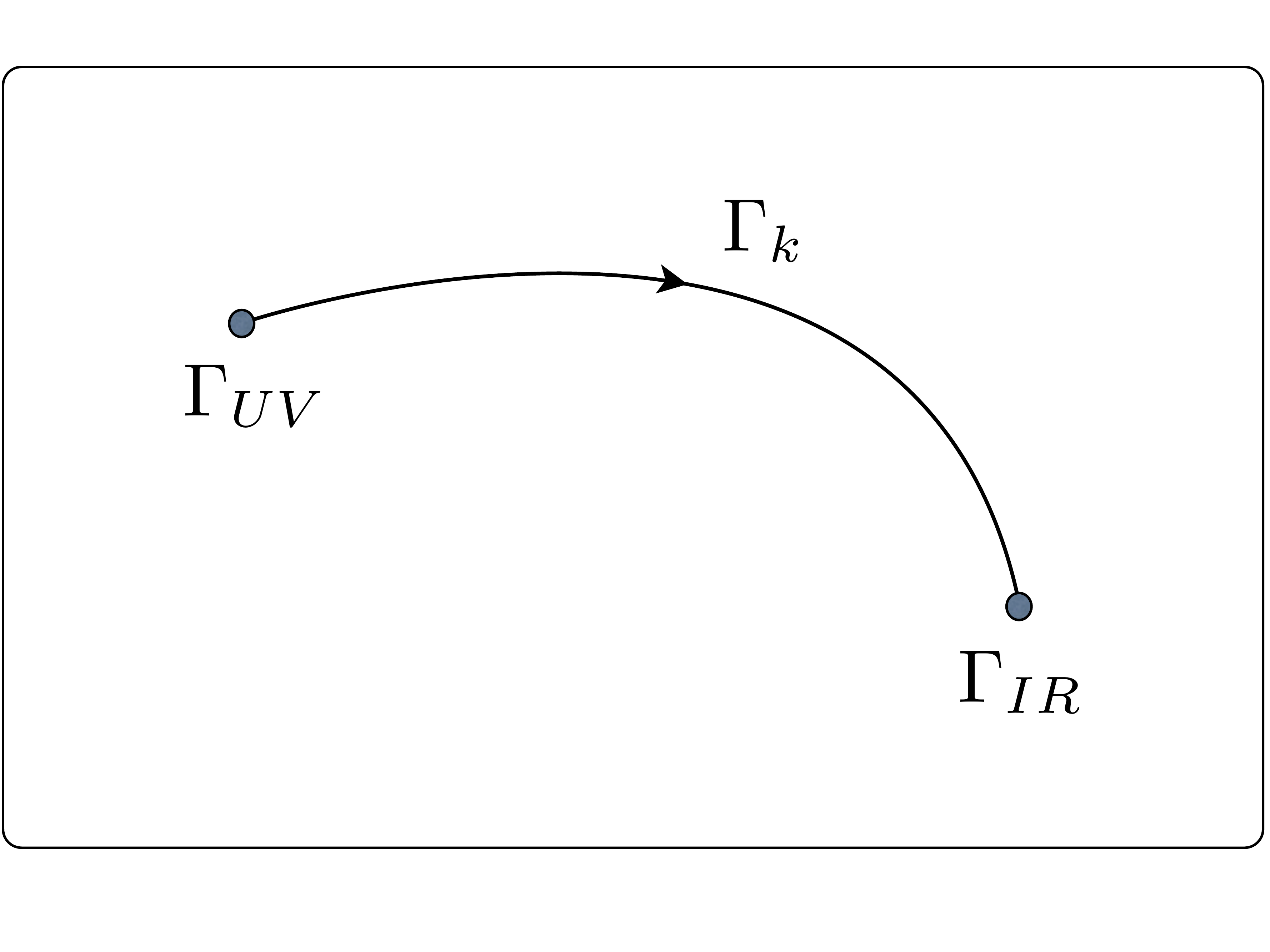} 
\caption{Cartoon depicting the flow in theory space: if the UV action $\Gamma_{UV}$ satisfies the Wess--Zumino relation (\ref{WZR}) with $c=0$, then the IR action $\Gamma_{IR}$ satisfies the Wess--Zumino relation (\ref{WZdC}) and the EAA, interpolating between the two, must satisfy (\ref{runningWZ}) with $\mathcal{C}_k=c_k-c_{UV}$.\label{fig: cartoon}}
\end{centering}
\end{figure}

In the functional RG framework, one considers an IR regulator quadratic in the fields:
\be
\Delta S_{k}\left[\phi,g\right]=\frac{1}{2}\int d^2x\sqrt{g}\phi R_{k}(\Delta)\phi\,,
\ee
chosen to suppress field modes in a covariant way: if $\phi_{n}$
is an eigenfunction of the covariant Laplacian $\Delta\phi_{n}=\lambda_{n}\phi_{n}$,
$R_{k}$ will act as a mass insertion for modes with $\lambda_{n}\ll k^{2}$,
while leaving unchanged the ones with $\lambda_{n}\gg k^{2}$. 
In this way we obtain a scale--dependent partition function:
\be
Z_k[J,g]=e^{W_{k}\left[J,g\right]}=\int{\cal D}\phi\, e^{-S\left[\phi,g\right]-\Delta S_{k}\left[\phi,g\right]+\int \sqrt{g}J\phi}\,.
\ee
The effective average action is then defined as the (shifted) Legendre transform:
\be
\Gamma_{k}[\varphi,g]=\int d^2x\sqrt{g}J_{\varphi}\varphi-W_{k}\left[J_{\varphi},g\right]-\Delta S_{k}\left[\varphi,g\right]\,,
\ee
where $\varphi=\left\langle \phi\right\rangle $ and $J_{\varphi}$
is obtained by inverting the solution of $\frac{\delta W_{k}\left[J,g\right]}{\delta J}=\varphi_{J}$.
By using its definition in the path integral, one finds the integro--differential
equation satisfied by the EAA (in which $\phi=\varphi+\chi$):
\be
e^{-\Gamma_{k}[\varphi,g]}=\int_{1PI} D\chi \,e^{-S_{UV}[\varphi+\chi,g]-c_{UV}S_{P}\left[g\right]-\Delta S_{k}[\chi,g]
}
\,.  \label{defEAA}
\ee
The main virtue of these definitions is that the EAA satisfies an exact RG flow equation \cite{Wetterich:1992yh}.
A scale derivative of (\ref{defEAA}) gives:
\be
\partial_{t}\Gamma_{k}[\varphi,g]=\left\langle \partial_{t}\Delta S_{k}[\chi,g]\right\rangle = \frac{1}{2}\textrm{Tr}\left\{\left\langle \chi_{A}\chi_{B}\right\rangle\partial_{t}R_{k}^{AB}[g]\right\}\,,
\label{preexactflow}
\ee
in which the expectation values are calculated with the fRG--regularized
path integral. Using the fact that the Legendre transform of the
generator of connected correlation functions is $\Gamma_{k}+\Delta S_{k}$,
we have:
\be
\left\langle \chi_{A}\chi_{B}\right\rangle = \left(\frac{\delta^{2}\Gamma_{k}[\varphi,g]}{\delta\varphi_{A}\delta\varphi_{B}}+R_{k}^{AB}[g]\right)^{-1}\,.
\ee
Substituting back in the previous expression we find the functional RG equation satisfied by the EAA:
\be
\partial_{t}\Gamma_{k}[\varphi,g]=\frac{1}{2}\textrm{Tr} \left(\frac{\delta^{2}\Gamma_{k}[\varphi,g]}{\delta\varphi\delta\varphi}+R_{k}[g]\right)^{-1}\partial_{t}R_{k}[g]\,.
\label{exactflow}
\ee
This equation is well defined, exact and offers a way to define QFTs non-perturbatively \cite{Berges:2000ew}.

From the exact flow equation for the EAA we obtain a corresponding equation for the $c$--function.
In particular, we can express the r.h.s. of (\ref{dtc}) using (\ref{preexactflow}):
\be
\partial_{t}c_{k}=-24\pi \,\partial_t \Gamma_k[e^{w \tau}\varphi,e^{2\tau}\delta]\Big|_{\int\tau \Delta \tau}\,.
\label{dtcexact}
\ee
Equation (\ref{dtcexact}) is the exact flow equation for the $c$--function in the fRG framework.
Using (\ref{exactflow}) in the r.h.s. leads to the following explicit form:
\be
\partial_{t}c_{k}  =  -12\pi\left. \textrm{Tr} \left(\frac{\partial_{t}R_{k}[\tau]}{\Gamma_{k}^{\left(2,0\right)}\left[\varphi,\tau\right]+R_{k}[\tau]}\right)\right|_{\int\tau\Delta\tau}
\,,
\label{dtcexact2}
\ee
where we defined $\Gamma[\varphi,\tau]\equiv\Gamma[e^{w \tau} \varphi,e^{2\tau}\delta]$ and $R_k[\tau]\equiv R_k[e^{2\tau}\delta]$.
The exact RG flow equation for the $c$--function is the main result of this section.

To write more explicitly the flow equation for the $c$--function we define the regularized propagator $G_k[\tau]\equiv (\Gamma^{(2,0)}[\varphi,\tau]+R_k[\tau])^{-1}$, perform two functional derivatives of (\ref{dtcexact}) with respect to the dilaton, set $\tau=0$ and extract the term proportional to $\Delta$:
\begin{eqnarray}
\partial_{t}c_{k}	&=&-24\pi\Big\{\textrm{Tr}\, G_{k}\left(\Gamma_{k}^{(2,1)}+R_{k}^{(1)}\right)G_{k}\left(\Gamma_{k}^{(2,1)}+R_{k}^{(1)}\right)G_{k}\partial_{t}R_{k}\nonumber\\
&&-\frac{1}{2}\textrm{Tr}\, G_{k}\left(\Gamma_{k}^{(2,2)}+R_{k}^{(2)}\right)G_{k}\partial_{t}R_{k}\nonumber\\
&&-\textrm{Tr}\, G_{k}\left(\Gamma_{k}^{(2,1)}+R_{k}^{(1)}\right)G_{k}\partial_{t}R_{k}^{(1)}+\frac{1}{2}\textrm{Tr}\, G_{k}\partial_{t}R_{k}^{(2)}\Big\}\Big|_{\Delta}\,,
\label{dtcexplict}
\end{eqnarray}
where all quantities are evaluated at $\varphi=\tau=0$.
Note that in (\ref{dtcexplict}) we had to derive the cutoff kernel $R_k$, since this depends explicitly on the dilaton.
As shown in \cite{Codello:2013wxa}, these additional terms in the flow equation for the proper--vertices are crucial in maintaining background symmetry when employing the background field method. 

\begin{figure}
\begin{centering}
\includegraphics[scale=.55]{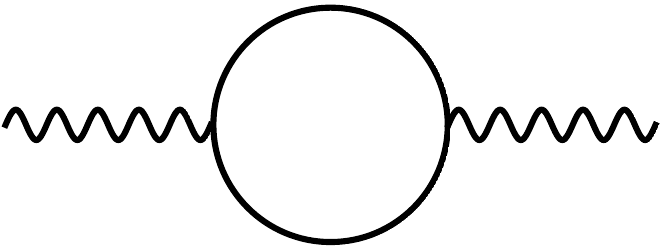}\qquad\qquad\qquad\includegraphics[scale=.7]{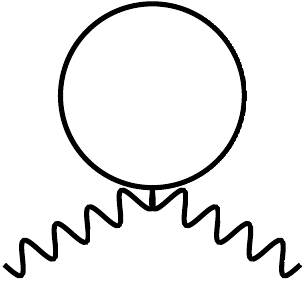}
\caption{Diagrammatic representation of the two terms in the r.h.s. of the flow equation (\ref{dtccompact}) for the $c$--function. \label{diagram_flow_eqn}}
\end{centering}
\end{figure}

The flow equation in the form (\ref{dtcexplict}) is a bit cumbersome so we introduce a compact notation to rewrite it in a simpler way.
If we introduce the formal operator $\tilde{\partial}_{t}=\partial_{t}R_{k}\frac{\partial}{\partial R_{k}}$, we can rewrite the flow equation (\ref{dtcexact}) for the $c$--function as:
\be
\partial_{t}c_k=-12\pi\,\textrm{Tr}\,\tilde{\partial}_{t}\log G_{k}[\tau]\Big|_{\int \tau \Delta \tau}\,,
\ee
where we used the following simple relations:
\begin{eqnarray}
\tilde{\partial}_{t}G_{k}[\tau]=-G_{k}[\tau]\partial_{t}R_{k}[\tau]G_{k}[\tau]\qquad\qquad\tilde{\partial}_{t}\log G_{k}[\tau]=G_{k}^{-1}[\tau]\tilde{\partial}_{t}G_{k}[\tau]=G_{k}[\tau]\partial_{t}R_{k}[\tau]\,.\nonumber
\label{relations}
\end{eqnarray}
Now we can rewrite the flow equation (\ref{dtcexplict}) in the following compact form:
\begin{eqnarray}
\partial_{t}c_{k}	  &=& 12\pi\,\textrm{Tr}\,\tilde{\partial}_{t}\left\{ \left(\Gamma_{k}^{(2,1)}+R_{k}^{(1)}\right)G_{k}\left(\Gamma_{k}^{(2,1)}+R_{k}^{(1)}\right)G_{k}\right\} \nonumber\\
		&&-12\pi\,\textrm{Tr}\,\tilde{\partial}_{t}\left\{ \left(\Gamma_{k}^{(2,2)}+R_{k}^{(2)}\right)G_{k}\right\}\Big|_\Delta\,,
\label{dtccompact}
\end{eqnarray}
where again all quantities are evaluated at $\varphi=\tau=0$.
This is the form that we will use in applications in section 5. Finally, we can represent diagrammatically the two terms on the r.h.s. of (\ref{dtccompact}) as in figure \ref{diagram_flow_eqn} and switch to momentum space to evaluate the diagrams by employing the techniques presented in \cite{Codello:2013wxa}.
In particular, continuous lines represent matter regularized propagators $G_k[0]$, while vertices with $m$--external wavy lines are the matter--dilaton vertices $\Gamma_k^{(2,m)}[\varphi,\tau]+R_k^{(m)}[\tau]$. Finally, each loop represents a $\int d^2x \,\tilde{\partial}_t$ or a $\int \frac{d^2q}{(2\pi)^2} \,\tilde{\partial}_t$ trace.
%
%
%
%
%

\subsection{fRG derivation of the integrated $c$--theorem}

We now rederive both the integrated $c$--theorem and the exact flow equation for the $c$--function using a fRG theory space perspective.

Away from a fixed point, 
apart for a Wess--Zumino term with running coefficient, that for the moment we call $\mathcal{C}_k$,
there must be many additional terms spoiling the fixed point Wess--Zumino relation (\ref{WZR}).
%
%
Since these terms vanish at a fixed point they must be proportional to the (dimensionless) beta functions. We can thus make the following ansatz:
\be
\Gamma_{k}[e^{w\tau}\varphi,e^{2\tau}g]-\Gamma_{k}[\varphi,g]={\cal C}_{k}\Gamma_{WZ}[\tau,g]+\beta\textrm{--terms}\,.
\label{runningWZ}
\ee
This relation can be read as a generalized running Wess--Zumino action.
The $\beta\textrm{--terms}$ indicate terms proportional to (at
least one) dimensionless beta function which vanish at the CFTs and are generated
along the flow by the fact that we are moving away from criticality.
%

If we now use the Weyl--invariant measure to construct the EAA, then 
at the UV fixed point, that is for $k\to\infty$, we must have
${\cal C}_{UV}=0$.
On the other hand, if we are not quantizing in a Weyl--invariant manner, we should reproduce the Wess--Zumino relation both at $k=\infty$ and $k=0$.
This tells us that in fact ${\cal C}_{k}=c_{k}-c_{UV}$ if the UV theory is quantized in a Weyl--invariant manner and ${\cal C}_{k}=c_{k}$ if not.
Weyl--invariant quantization corresponds, in the EAA formalism, to a constant shift of $\mathcal{C}_k$.

Relation (\ref{runningWZ}) can be used to give an equivalent RG derivation of the integrated $c$--theorem.
When we flow for $k\to0$ to a fixed point theory CFT$_{IR}$, relation (\ref{runningWZ}) tells us that:
\be
\Gamma_{IR}[e^{w\tau}\varphi,e^{2\tau}g]-\Gamma_{IR}[\varphi,g]=\left(c_{IR}-c_{UV}\right)\Gamma_{WZ}[\tau,g]\,.
\label{WZdC}
\ee
If we now set $g_{\mu\nu}=\delta_{\mu\nu}$ and expand $\Gamma_{IR}[\varphi,\tau] \equiv \Gamma_{IR}[e^{w\tau}\varphi,e^{2\tau}\delta]$ in powers of the dilaton we find:      
\begin{eqnarray}
\Gamma_{IR}[e^{w\tau}\varphi,e^{2\tau}\delta] & = & \Gamma_{IR}[\varphi,0]+\int d^2x\,\tau_{x}\frac{\delta}{\delta\tau_{x}}\Gamma_{IR}[\varphi,\tau]\Big|_{\tau\to0}\nonumber\\
 &  & +\frac{1}{2}\int d^2x \int d^2y\, \tau_{x}\tau_{y}\frac{\delta^{2}}{\delta\tau_{x}\delta\tau_{y}}\Gamma_{IR}[\varphi,\tau]\Big|_{\tau\to0}+\mathcal{O}(\tau^{3})\,.
\label{EA_Taylor}
\end{eqnarray}
The functional derivatives of the effective action are related to the traces of the energy--momentum tensor:				
\be
\left\langle \Theta_x\right\rangle_{IR}=\frac{\delta}{\delta\tau_{x}}\Gamma_{IR}[\varphi,\tau]\Big|_{\tau\to0}\qquad\qquad
%
\left\langle \Theta_x\Theta_y\right\rangle_{IR}=\frac{\delta^{2}}{\delta\tau_{x}\delta\tau_{y}}\Gamma_{IR}[\varphi,\tau]\Big|_{\tau\to0}\,.
\ee
The first relation is identically zero at a CFT, i.e. $\langle \Theta_x \rangle_{IR}=0\,$. 
Inserting the second relation in (\ref{EA_Taylor}) and expanding, as before, $\tau_y$ around $\tau_x$ using (\ref{tauexp}) gives immediately the integrated $c$--theorem (\ref{ct}).
This derivation represents a consistency of the ansatz (\ref{runningWZ}).
%

It is now clear that from the Wess--Zumino relation at finite $k$ (\ref{runningWZ}) we can easily read
off the flow of the central charge.
In this way, since  $\partial_{t}c_{k}=\partial_{t}\mathcal{C}_{k}$,
from the coefficient of $\int\tau\Delta\tau$ in $\partial_{t}\Gamma_{k}[e^{w\tau}\varphi,e^{2\tau}\delta]$
we recover the exact RG flow equation for the $c$--function (\ref{dtcexact}).

Another way to see that the flow of the $c$--function is given by (\ref{dtcexact}) is to recognize that $\mathcal{C}_k$ is nothing more than the coupling constant of the Polyakov action.
As we said, when working on curved backgrounds one should always add the Polyakov term to a truncation.
Thus the Wess--Zumino action on the r.h.s. of (\ref{runningWZ}) derives from the presence of the Polyakov action, with coefficient $\mathcal{C}_k$, in the EAAs on the l.h.s of the same equation.
Then, as just seen in the previous paragraph, a $t$--derivative relates $\partial_t c_k$  to the two--point function of the dilaton. 
In principle one can obtain the flow of $\mathcal{C}_k$ directly as the coefficient of $\int \sqrt{g} R \frac{1}{\Delta}R$ but this is more laborious. 
%
%
Finally, note that the inclusion of the Polyakov action with running central charge makes the truncation consistent with the conformal anomaly both in the UV and in the IR. To understand the $\beta$--terms we will consider, in the next section, the scale anomaly.

\section{General form of the effective average action}

In this section we put forward some requirements which an ansatz for the EAA should satisfy. These requirements are motivated from the fact that the EAA should reproduce some generic features of QFTs, namely the scale and the conformal anomaly. In particular we will try to shed light on the nature of the $\beta$--terms introduced in equation (\ref{runningWZ}).

\subsection{The local ansatz and its limitations}

When studying truncations of the EAA, one generally starts by expanding the functional in terms of local operators compatible with the symmetries of the system:
\be
\Gamma_k[\varphi,g]=\sum_{i}g_{i,k}\int d^2x\,\sqrt{g}\,\mathcal{O}_i[\varphi,g]\,.
\label{locEAA}
\ee
This equation defines the running coupling constants $g_{i,k}$, which become the coordinates that parametrize theory space in the given operator basis. 

A class of operators, which is not complete, but allows many computations to be performed analytically,
is the one composed of powers of the field, i.e. $\mathcal{O}_i[\varphi,g]= \varphi^{2i}$ and $g_{i,k}=\frac{\lambda_{2i,k}}{(2i)!}$.
In this approximation, one usually re--sums the field powers into a running effective potential $V_k(\varphi)$ and equivalently considers the following ansatz for the EAA:
\be
\Gamma_k[\varphi,g]=\int d^2x\sqrt{g}\left[   \frac{1}{2}\varphi \Delta \varphi +V_k(\varphi)\right]\,,
\label{LPA}
\ee
known as local potential approximation (LPA).
Within this truncation the exact flow equation (\ref{exactflow}) becomes a partial differential equation:
\be
\partial_t V_k(\varphi)=c_d \frac{k^d}{1+V_k''(\varphi)/k^2}\,,
\label{dtLPA}
\ee
with $c_d^{-1} = (4\pi)^{d/2} \Gamma(d/2+1)$.
Even such a simple truncation is able to manifest qualitatively all the critical information relative to the theory space of scalar theories and in particular the fixed point structure \cite{Codello:2012sc}.

However, the effective action usually contains also nonlocal terms.
%
Some of these nonlocal terms are directly related to the finite part of the effective action \cite{Codello:2010mj},
which generally has a complicated form encoding all the information contained in the correlation functions or amplitudes.
These terms are not present in the LPA which can be seen as the limit where we discard all the momentum structure of the vertices.

Nevertheless there are other nonlocal terms that are non--zero only away from a fixed point: these are the $\beta$--terms introduced in equation (\ref{runningWZ}).
As we will explain in this section these terms are needed to recover known results and will play a central role in our computations.
If we limit ourselves to the local truncation ansatz (\ref{locEAA}),
then one finds that the flow equation for the $c$--function
is driven only by the classical non Weyl--invariant terms, which is not correct.
This is not due to the fact that the flow equation (\ref{dtcexact}) is wrong, rather, it is the truncation ansatz (\ref{locEAA}) that is insufficient.
Fluctuations induce the $\beta$--terms of equation (\ref{runningWZ}) and we will see that these are crucial in driving the flow of the $c$--function.

We will argue that these nonlocal terms have a precise form. We will do this requiring the EAA to reproduce the scale anomaly.

\subsection{Nonlocal ansatz and  the scale anomaly}

It is easy to understand the origin of the terms on the r.h.s. of (\ref{runningWZ}) which are linear in $\tau$: they are related to the scale anomaly.
To see this let us rescale the fields and expand the EAA in powers of the dilaton:
\be
\Gamma_k[\varphi,\tau] = \Gamma_k[\varphi,0] + \int d^2x\, \tau \, \langle \Theta\rangle_k + O(\tau^2)\,,
\label{linearterms}
\ee
where:
\be
\langle \Theta \rangle_k
=\frac{\delta}{\delta \tau}\Gamma_{k}[\varphi, \tau] \Big|_{\tau\to0}
\,,
\ee
defines the scale dependent energy--momentum tensor trace.
In the IR the EAA reduces to the standard effective action, which generally is scale anomalous. 
If we start with some UV action deformed by terms of the form $\sum_{j}g_{j}\int d^2x\,\sqrt{g}\,\mathcal{O}_i$, the corresponding scale anomaly in flat space reads:
\be
\int d^2x\,\sqrt{g}\left\langle \Theta\right\rangle_{IR} =-\sum_{i}\left(\beta_{i}-d_ig_{i}\right)\int d^2x\,\mathcal{O}_i[\varphi,\delta] \,,
\ee
where $d_i$ are the dimensions of the coupling constants. 
The expression in brackets is nothing but the beta function of the dimensionless coupling:
\be
k^{d_i}\tilde{\beta}_{i}=\beta_{i} -d_i g_{i}\,.
\label{dimlessbeta}
\ee
This is a standard result known from both ordinary and conformal perturbation theories \cite{mussardo}.

Now we consider again the $\beta$--terms on the r.h.s. of (\ref{runningWZ}). They come from the conformal variation of the EAA which should include also the terms due to the scale anomaly. Therefore it is natural to generalize the above equation for a generic $k$:
\be
\langle \Theta_x\rangle_k=-\sum_{i}k^{d_i}\tilde{\beta}_{i} \int d^2x\,\mathcal{O}_i[\varphi,\delta]\,.
\ee
If we insert this into (\ref{linearterms}) we find:
\be
\Gamma_k[\varphi,\tau] = \Gamma_k[\varphi,0] -\tau\sum_{i}k^{d_i}\tilde{\beta}_{i}\int d^2x\,\mathcal{O}_i[\varphi,\delta] + O(\tau^2)\,.
\label{lintau}
\ee
This expression gives a non trivial flow of the $c$--function since we now have the vertex
%
\be
\left.\Gamma_{k}^{(2,1)}[\varphi,\tau]\right|_{\varphi=\tau=0}=-\sum_{i}k^{d_i}\tilde{\beta}_{i}\int \mathcal{O}_i^{(2,0)}[0,0]
\ee
to insert in the r.h.s. of the exact flow equation (\ref{dtccompact}).

We now propose a covariant form for (\ref{lintau}) using the following properties:
\be
\qquad g_{\mu\nu} \rightarrow  e^{2\tau} g_{\mu\nu}\qquad\to\qquad\frac{1}{2\Delta} R \rightarrow \frac{1}{2\Delta} R +\tau \,.
\ee 
With this and ${\cal O}_i \rightarrow e^{w_i \tau} {\cal O}_i$, it is easy to verify that the action
\be
\Gamma_k[\varphi,g] = 
\sum_{i}g_{i,k}\int\sqrt{g}\,\mathcal{O}_i[\varphi,g]
-\frac{1}{2}\sum_{i}\beta_{i}\int\sqrt{g}\,\mathcal{O}_i[\varphi,g]\frac{1}{\Delta}R + \cdots \,,
\label{nonlocEAA0}
\ee
reproduces (\ref{lintau}) to linear order in $\tau$.
In order to get an  ansatz consistent also with the conformal anomaly we need to add to
(\ref{nonlocEAA0}) the Polyakov term with as coefficient the running central charge $\mathcal{C}_k$:
\be
\Gamma_k[\varphi,g] =  \sum_{i}g_{i,k}\int\sqrt{g}\,\mathcal{O}_i[\varphi,g] 
-\frac{1}{2}\sum_{i}\beta_{i}\int\sqrt{g}\,\mathcal{O}_i[\varphi,g]\frac{1}{\Delta}R
-\frac{\mathcal{C}_k}{96\pi}\int\sqrt{g} \, R \frac{1}{\Delta}R \,.
\label{nonlocEAA}
\ee
The form (\ref{nonlocEAA}) represents a parametrization of the EAA consistent with (\ref{runningWZ}) to linear order in the beta functions and hints to what could be the general for of the EAA away from criticality. 
For the time being we will not improve further our ansatz, since we will see in the next section that the understanding of the linear terms in the beta functions is already sufficient to build the $c$--function in some non--trivial cases.
%
%
We hope to come back to the issue of higher order terms in $\tau$, which may play a role in making a bridge between the fRG perspective adopted here and the ideas related to the local RG \cite{Osborn:1991gm}.

\section{Applications}

\subsection{Checking exact results}

Here we provide two examples where the $c$--function
and the difference $c_{UV}-c_{IR}$ are computed and can be compared
to known exact results.
We will consider a free scalar
field and a free (Majorana) fermionic field whose fixed point actions are perturbed
by a mass term, so they flow to $c_{IR}=0$.
%
%

\subsubsection{Massive deformation of the Gaussian fixed point}

\begin{figure}
\begin{centering}
\includegraphics[scale=0.45]{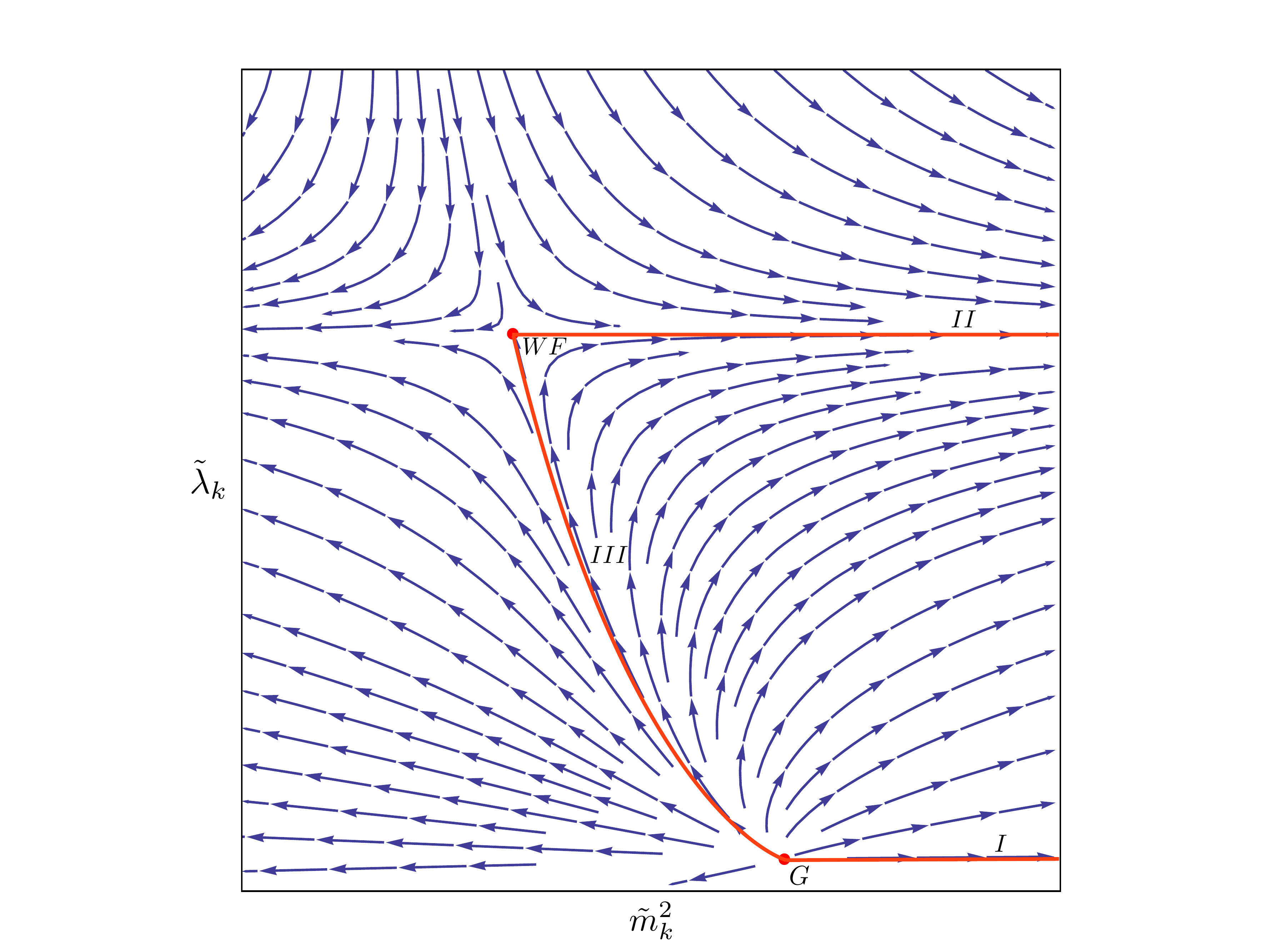} 
\caption{The flow in the $(\tilde{m}^2_k,\tilde{\lambda}_k)$ plane showing the Gaussian (G) and Ising (WF) fixed points. The flow induced by the massive deformation of the Gaussian fixed point is represented by trajectory--$I$, the flow induced by the massive deformation of the Ising fixed point is represented by trajectory--$II$ while the flow between the two fixed points happens along trajectory--$III$.\label{fig: plot_flow}}
\end{centering}
\end{figure}

We consider a scalar field with Gaussian action and $c_{UV}=1$ perturbed by a mass term.
Since the beta function of the mass is zero (there are no interactions),
our general ansatz (\ref{nonlocEAA}) for the EAA reads:
\be
\Gamma_{k}[\phi,g] = \frac{1}{2}\int d^2x \sqrt{g} \, \phi(\Delta+m^{2})\phi-\frac{c_{k}}{96\pi}\int\sqrt{g}R\frac{1}{\Delta}R\,,
\label{scalar}
\ee
or when we rescale the fields:
\be
\Gamma_{k}[\phi,e^{2\tau}\delta] = \frac{1}{2}\int d^2x\,\phi\left(\Delta+e^{2\tau}m^{2}\right)\phi-\frac{c_{k}}{24\pi}\int\tau\Delta\tau\,.
\ee
It's clear that the only interaction between $\phi$ and $\tau$ is the one induced by the dimension of the mass.
In order to avoid possible vertices coming from the cutoff action
we use the mass cutoff $R_{k}(z)=ak^{2}$ which has the advantage
of having no dependence with respect to the background metric.
We have introduced the parameter $a$ to check the cutoff independence of the result.
After a short computation\footnote{We need to evaluate the first diagram of figure \ref{diagram_flow_eqn}, for more details see section 5.2.} we find the following flow:
\be
\partial_{t}c_{k} =
\frac{4 a k^2 m^4}{\left(a k^2+m^2 \right)^3}\,,
\label{c_scalar}
\ee
where $m$ is the dimensionful mass.                
This RG flow occurs along trajectory--$I$ of figure \ref{fig: plot_flow}.

Integrating the above differential equation, with the initial condition
$c_{\infty}=1$ (the central charge of the Gaussian fixed point) we find:
\be
c_{k}=1-\frac{m^{4}}{\left(a k^{2}+m^{2}\right)^{2}}\,.
\ee
In the $k\rightarrow 0$ limit this gives $c_{0}=0$ which implies $\Delta c=1$ independently of the cutoff  parameter $a$.
As expected a massive deformation of the Gaussian fixed point leads in the IR to a theory with zero central charge.
%

\subsubsection{Massive deformation of the Ising fixed point}

In this example we make a massive deformation of the Ising fixed point.
The critical Ising model is described by a free Majorana fermion and a massive deformation of this corresponds to consider $T>T_c$ \cite{mussardo}. 
According to our general ansatz (\ref{nonlocEAA}) and considering that, as before, the mass
beta function is zero, the EAA reads:
\be
\Gamma_{k}[\bar{\psi},\psi,g] = \int d^2x\sqrt{g}\, \bar{\psi}\left(\slashed{\nabla}+m\right)\psi-\frac{c_{k}}{96\pi}\int\sqrt{g}R\frac{1}{\Delta}R\,,
\ee
or after the rescaling:
\be
\Gamma_{k}[e^{\tau/2} \bar{\psi},e^{\tau/2}\psi,e^{2\tau}\delta] =\int d^2x\,\bar{\psi}\left(\slashed{\nabla}+e^{\tau}m\right)\psi-\frac{c_{k}}{24\pi}\int\tau\Delta\tau\,.
\ee
The computation proceeds along the lines of the scalar case.
Once again we use the mass cutoff $R_{k}=ak$ and we find:
\begin{eqnarray*}
\partial_{t}c_{k} 
=\frac{akm^{2}}{\left(ak+m\right)^3}.
\end{eqnarray*}
This RG flow occurs along trajectory--$II$ of figure \ref{fig: plot_flow}.

Integrating this equation with boundary condition $c_{\infty}=\frac{1}{2}$ (the central charge of the Ising model)
leads to
\be
c_{k}=\frac{1}{2}-\frac{m^2}{2\left(ak+m\right)^{2}}\,,
\ee
which gives $c_{0}=0$ and $\Delta c=\frac{1}{2}$ as expected.
%

\subsection{The $c$--function in the local potential approximation}

The local potential approximation (LPA), introduced in section 4.1, is characterized by the action  (\ref{LPA}); 
%
%
in our case generalizes to:
\be
\Gamma_{k}[\varphi,g] =  \int d^2x \sqrt{g}\left[\frac{1}{2}\varphi\Delta\varphi+V_{k}(\varphi)-\frac{1}{2}\partial_{t}V_{k}(\varphi)\frac{1}{\Delta}R
-\frac{c_{k}}{96\pi}R\frac{1}{\Delta}R\right]\,,
\label{LPAansatz0}
\ee
or after rescaling the fields:
\be
\Gamma_k [\varphi,e^{2\tau}\delta] = \int d^2x \left[\frac{1}{2}\varphi\Delta\varphi+e^{2\tau} V_k(\varphi)-\partial_{t}V_k(\varphi)\,\tau
-\frac{c_{k}}{24\pi} \tau\Delta\tau \right] \,.
\label{LPAansatz}
\ee
If we now pass to dimensionless variables, $\varphi=k^{- w}\tilde{\varphi}$ and $V_{k}(\varphi)=k^{2}\tilde{V}_{k}(\tilde{\varphi})$,
then the second and third terms in the above equation, to linear order in $\tau$,
become $V_k(\varphi)-k^2 \partial_t \tilde{V}_k(\tilde{\varphi}) \tau$, so that
the scalar--dilaton interaction is proportional to the dimensionless scale derivative of the potential. 

To obtain the flow equation for the $c$--function we use (\ref{dtccompact}) and the mass cutoff $R_k(z)=k^2$ so that all cutoff vertices drop out.
Only the first diagram of figure \ref{diagram_flow_eqn} contributes terms of order $p^2$ in the external momenta,
more specifically we need to evaluate the integral:
\be
\partial_{t}c_{k}=-12\pi(\partial_{t}\tilde{V}_{k}^{\prime\prime}(\varphi_0))^{2}k^{4}\int\frac{d^2q}{(2\pi)^d}\,G_{k}^{2}(q^2)G_k\left((p+q)^2\right)\partial_{t}R_{k}(q^2)\Big|_{p^2}\,,
\label{c_LPA_0}
\ee
with the following regularized propagator:
\be
G_k(q^2)=\frac{1}{q^2+V''_k(\varphi_0)+R_k(q^2)}\,.
\ee
Here $\varphi_0$ is the minimum of the running effective potential, i.e. the solution of $V'_k(\varphi)=0$.
With the mass cutoff one finds the following result:
\be
\int\frac{d^2q}{(2\pi)^d}\,G_{k}^{2}(q^2)G_k\left((p+q)^2\right)\partial_{t}R_{k}(q^2)\Big|_{p^2} = -\frac{1}{12 \pi k^{4} (1+\tilde{V}_{k}^{\prime\prime}(\varphi_0)  )^3}\,,
\ee
provided that $\tilde{V}_{k}^{\prime\prime}(\varphi_0)>-1$, since otherwise the momentum integral does not converge.
Inserting this back in (\ref{c_LPA_0}) finally gives:
\be
\partial_{t}c_{k}=\frac{(\partial_{t}\tilde{V}_{k}^{\prime\prime}(\varphi_0))^{2}}{(1+\tilde{V}_{k}^{\prime\prime}(\varphi_0)  )^3}\,,
\label{c_LPA}
\ee
which is the flow equation for the $c$--function in the LPA with a mass cutoff. This the main result of this section.
Note that since (\ref{c_LPA}) is valid under the condition $\tilde{V}_{k}^{\prime\prime}(\varphi_0)>-1$, the $c$--theorem $\partial_{t}c_{k}\geq0$ is indeed satisfied within the LPA. 
All the beta functions in this section are computed from the two-point function of the running effective potential.
\begin{figure}
\begin{centering}
\includegraphics[scale=0.6]{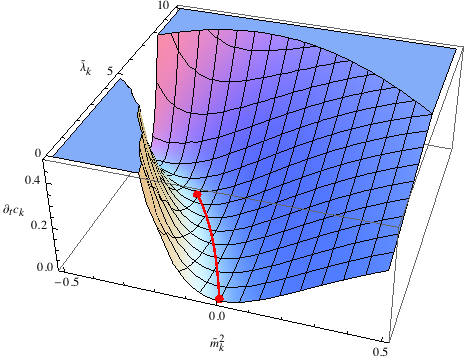} 
\caption{$\partial_t c_k$ in the $(\tilde{m}_k^2,\lambda_k)$ plane. We marked with a red dot the position of the Gaussian and Ising fixed points.
\label{fig: plot_c_flow}}
\end{centering}
\end{figure}
\subsubsection{Flow between the Gaussian and Ising fixed points}

We now consider the simple case where there are just two running couplings parametrizing theory space, i.e. we expand the running effective potential in a Taylor series:
\be
V_{k}(\varphi)=\frac{1}{2!}m_{k}^{2}\varphi^{2}+\frac{1}{4!}\lambda_{k}\varphi^{4}+...
\label{LPA_Taylor}
\ee
where $m_k^2$ is the mass and $\lambda_k$ the quartic self--interaction.
Inserting (\ref{LPA_Taylor}) in the flow equation for the effective potential (\ref{dtLPA}) and projecting out the flow of the two couplings gives, after passing to dimensionless variables $m_{k}^{2}=k^{2}\tilde{m}_{k}^{2}$ and $\lambda_{k}^{2}=k^{2}\tilde{\lambda}_{k}$, the following system of beta functions:
%
\be
\partial_{t}\tilde{m}_{k}^{2}  =  -2\tilde{m}_{k}^{2}-\frac{1}{4\pi}\frac{\tilde{\lambda}_{k}}{(1+\tilde{m}_{k}^{2})}\qquad\qquad
\partial_{t}\tilde{\lambda}_{k}  =  -2\tilde{\lambda}_{k}+\frac{3}{2\pi}\frac{\tilde{\lambda}_{k}^{2}}{(1+\tilde{m}_{k}^{2})^{2}}\,.
\label{LPA_bf}
\ee
This system has two fixed points: the Gaussian $(\tilde{m}_k^2,\tilde{\lambda}_k)=(0,0)$ and the Ising $(\tilde{m}_k^2,\tilde{\lambda}_k)=(-\frac{1}{4},\frac{3\pi}{2})$.
The Gaussian fixed point has two IR repulsive directions, while the Ising fixed point has one IR repulsive and one IR attractive direction.
The trajectories starting along these directions are shown in figure \ref{fig: plot_flow}, in particular trajectory--$III$ connects the two fixed points.

We can now use (\ref{c_LPA}) to evaluate the $c$--function in this truncation.
This turns out to be simply related to the square of the dimensionless mass beta function:
%
\be
\partial_{t}c_{k}  =  \frac{1}{(1+\tilde{m}_{k}^{2})^3} \left( \partial_t \tilde{m}_k^2 \right)^2
=  \frac{1}{(1+\tilde{m}_{k}^{2})^{3}}\left(2\tilde{m}_{k}^{2}+\frac{1}{4\pi}\frac{\tilde{\lambda}_{k}}{(1+\tilde{m}_{k}^{2})^{2}}\right)^{2} \,.
\label{c_LPA_Taylor}
\ee
%
As for (\ref{c_LPA}), the result is only valid for $\tilde{m}_k^2>-1$, so in this range we do have
$\partial_{t}c_{k}\geq0$, which is consistent with the $c$--theorem.
The flow (\ref{c_LPA_Taylor}) is  similar to the one given in \cite{Haagensen:1993by}, which was there found "by trial and error".
%
Equation (\ref{c_LPA_Taylor}) is the first non--trivial example of explicit flow equation for the $c$--function obtained using the procedure presented in this work.
In figure \ref{fig: plot_c_flow} we plot $\partial_t c_k$ in the plane $(\tilde{m}_k^2,\tilde{\lambda}_k)$:
one can see that the magnitude of $\partial_t c_k$ is smaller along a "valley" containing the two fixed points.
Along this valley lies the trajectory connecting them, trajectory--$III$ of figure \ref{fig: plot_flow}.

We computed $\Delta c$ by integrating the flow of the central charge along the path connecting the Gaussian and Ising fixed points,
but in this simple truncation there is not quantitative agreement, namely the difference between the two central charges is very small. This is due to the fact that along this trajectory the mass beta function is also very small. To improve our result we need to consider a more refined truncation ansatz for the running effective potential.
%
We leave these studies to future work.

\subsubsection{Sine--Gordon model}

We now consider the Sine--Gordon model which, in the continuum limit, is described by the following action \cite{mussardo}:
\be
S_{SG}[\phi]=\int d^2x\left[\frac{1}{2} \phi \Delta \phi-\frac{m^2}{\beta^{2}}\left(\cos\left(\beta\phi\right)-1\right) \right]\,,
\ee
where $m$ is the mass and $\beta$ is a coupling constant.
This theory can be seen as a massive deformation of the Gaussian fixed point action
(with $c_{UV}=1$)
%
%
and indeed we will find $c_{IR}=0$.

\begin{figure}
\begin{centering}
\includegraphics[scale=0.25]{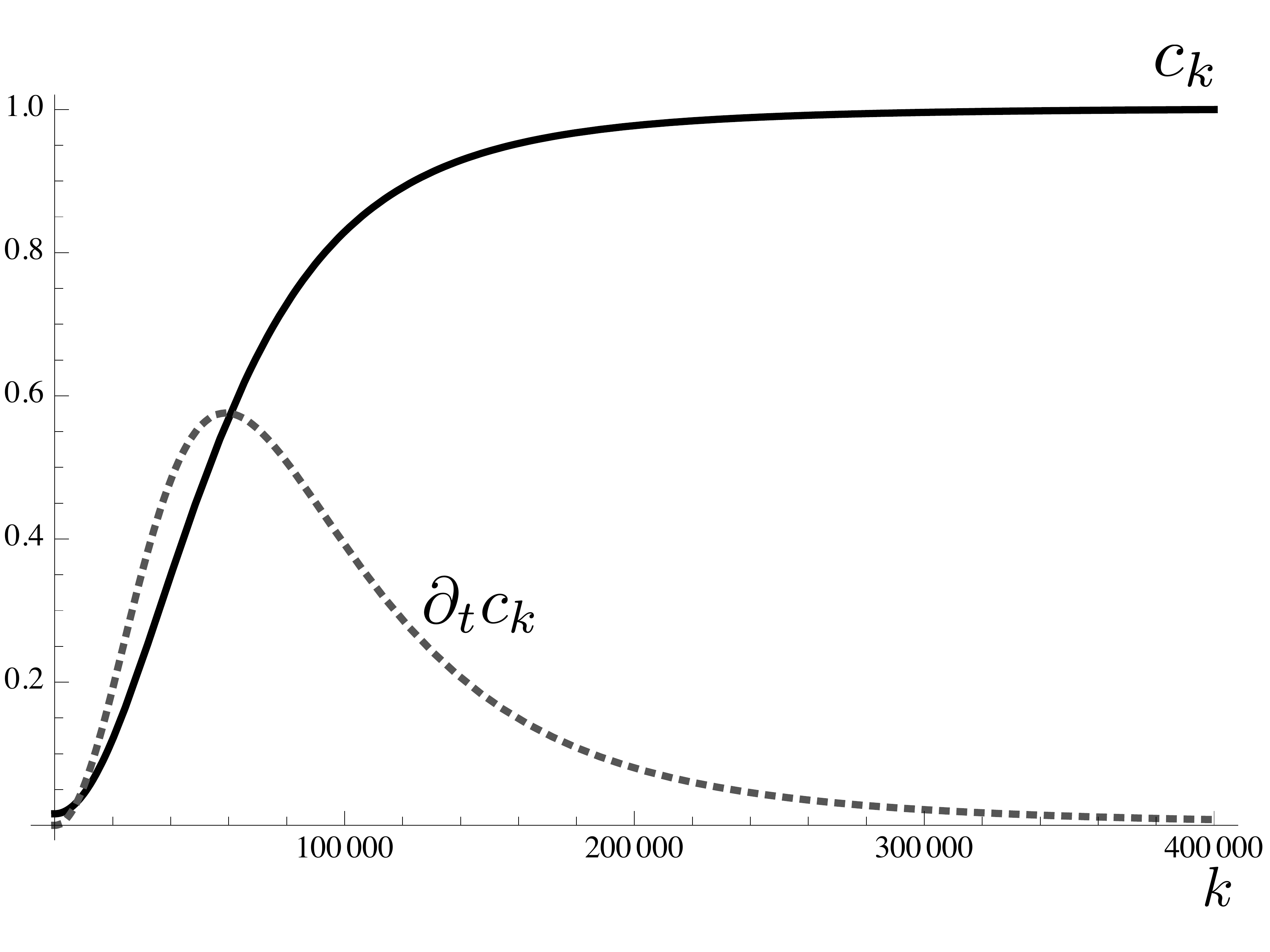} 
\caption{Flow of the Sine-Gordon model: the continuos line shows the running of the $c$-function and the dotted line has a bell shape meaning that the beta function of $c_k$ is zero at the endpoints of the flow.\label{fig: plot_SG}}
\end{centering}
\end{figure}

The Sine-Gordon model can be described by an LPA with effective potential:
\be
V_k(\varphi) = -\frac{m_{k}^2}{\beta_{k}^{2}}\left(\cos\left(\beta_{k}\varphi\right)-1\right)\,.
\label{pot_SG}
\ee
%
%
%
%
We find the following form for the beta functions of $m_{k}$ and $\beta_{k}$:
\footnote{The running of the couplings is found via the expansion of the two-point function of the running effective potential. This is not the best procedure since it would be more natural to perform a Fourier expansion. See also \cite{Nagy:2009pj}.}
\begin{eqnarray*}
\partial_{t}\tilde{m}_{k}^2 & = & \frac{\tilde{m}^2_k\left(\beta^{2}_k-8\pi\left(1+\tilde{m}^2_k\right)\right)}{4\pi\left(1+\tilde{m}^2_k\right)}\\
\partial_{t}\beta_{k} & = & -\frac{3\tilde{m}^2_k\beta^{3}_k}{8\pi\left(1+\tilde{m}^2_k\right)^{2}}\,,
\end{eqnarray*}
where $\tilde{m}_{k}^2=m_{k}^2/ k^{2}$ is the dimensionless mass.
Inserting the Sine--Gordon running potential (\ref{pot_SG}) into the flow equation (\ref{c_LPA}) now gives:
\be
\partial_{t}c_{k}=\frac{\widetilde{m}_k^{4}\left(\beta^{2}_k-8\pi\left(1+\tilde{m}_k^2\right)\right)^{2}}{16\pi^{2}\left(1+\tilde{m}_k^2\right)^{5}}\,.
\ee
We solved the system of equations numerically imposing $c_{UV}=1$ finding 
$\Delta c\simeq0.998$, in satisfactory
agreement with the exact result $\Delta c=1$.
%
%
In figure \ref{fig: plot_SG} we plot the running of $c_k$ as well as its beta function.

\subsection{The $c$--function in the loop expansion}

The last approximation we will consider is the loop expansion.
The exact flow equation for the EAA (\ref{exactflow}) can be solved perturbatively \cite{Litim:2001ky,Codello:2013bra} loop by loop.
We review this in the Appendix.
%
In the first part of this section we will look at the various contributions diagrammatically, while in the second part we will explicitly evaluate one subclass of these.
%

\subsubsection{Zamolodchikov's metric: diagrammatics}

Using relation (\ref{loopgen}) we can compute the running of the EAA
at each order in the loop expansion.
The running of the $L$--th term $\partial_{t}\Gamma_{L,k}$,
say, will contain a contribution to the running of $c_{k}$ that we will call $\partial_{t}c_{L,k}$.
%
%
The term $c_{L,k}$ 
arises only from diagrams with $L$ matter loops and two dilaton
external lines.
In this way we can build a loop expansion for the
$c$--function.

We can start by applying this construction step by step so to make
clear how everything works.
We will work with a $\mathbb{Z}_{2}$--symmetric
scalar theory, so that
%
the part linear in the dilaton of our general ansatz (\ref{nonlocEAA}) takes the form:
\be
\sum_{n}\frac{1}{(2n)!}\tilde{\beta}_{2n}\varphi^{2n}\tau\,,
\label{vert}
\ee
where $\tilde{\beta}_{2}$ is the mass beta function, $\tilde{\beta}_{4}$ is the $\varphi^{4}$ coupling beta function, and so on.

At one loop, we have only the following diagram, obtained from (\ref{loop1}) of the Appendix by functional derivation with respect to the dilaton,
\begin{center}
\includegraphics[scale=.5]{1loopb.pdf}
\end{center}
Here we adopt the same diagrammatic rules of section 3.2 where the continuous line represents the regularized propagator
(in this case given in equation (\ref{prop}) of the Appendix), while the wavy line represents the dilaton.
On every diagram the operator $\tilde{\partial}_t$ acts, but in this case it is just $\partial_t$. 
In this diagram the vertices, as derived from (\ref{vert}), are the mass beta function, so this contribution goes
like $\tilde{\beta}_{2}^2$ and we recover the LPA result (\ref{c_LPA})
as one would expect.
%

From the flow of the two--loop contribution, (\ref{loop2}) of the Appendix, we obtain different terms.
We get the {}``non--diagonal'' contribution (we will make this jargon clear in a second):
\begin{center}
\includegraphics[scale=.5]{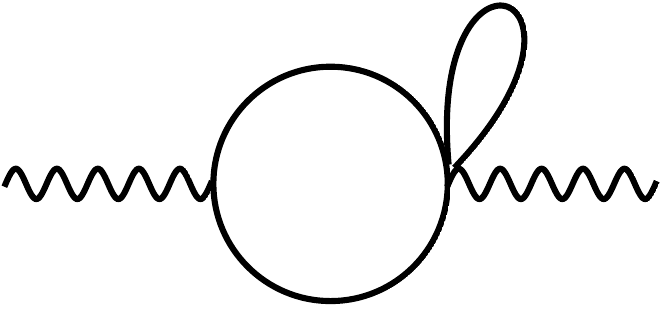}
\end{center}
proportional to $\tilde{\beta}_{2} \,\tilde{\beta}_{4}$.
%
Together with this, we also have the following 2--loop diagonal contributions:
\begin{center}
\includegraphics[scale=.6]{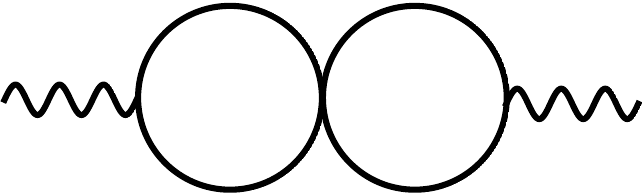}\qquad\qquad \includegraphics[angle=180,origin=c,scale=.6]{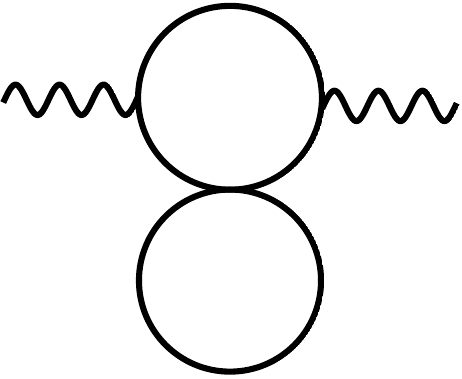}
\end{center}
which are proportional to $\lambda_{2}\,\tilde{\beta}_{2}^{2}$.
These represent a diagonal but coupling--dependent contribution, in the sense that couplings do not only appear through the beta functions. 
When going to 3--loops, 4--loops and so on, corresponding diagrams must be considered for all the diagonal contributions. 

At three loops (remember we are considering a $\mathbb{Z}_{2}$--symmetric
theory, so there are no scalar odd power interactions) we get again
the {}``diagonal'' contributions:
\begin{center}
\includegraphics[scale=.5,]{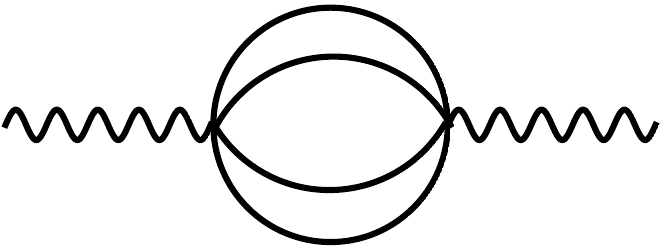}\qquad\qquad \includegraphics[scale=.5]{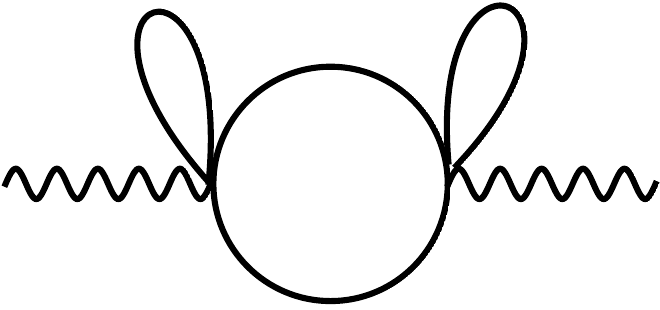} 
\end{center}
both proportional to $\tilde{\beta}_{4}^{2}$,
as well as a nondiagonal one:
\begin{center}
\includegraphics[scale=.5]{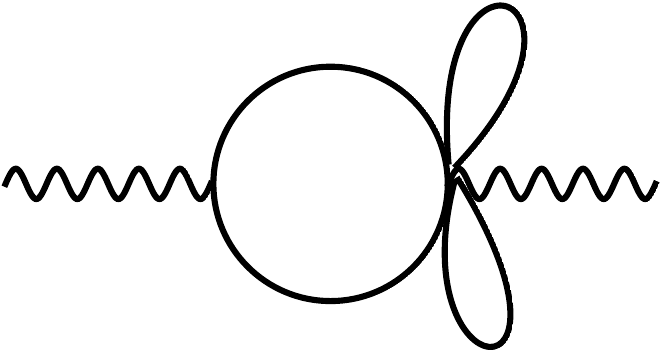}
\end{center}
proportional to $\tilde{\beta}_{2}\,\tilde{\beta}_{6}$.
%
%
From these first diagrams we clearly see that from the structure of the loop expansion we only get terms
quadratic in the beta functions.

We can indeed follow Zamolodchikov and define the {}``metric'' $g_{ij}$
through:
\be
\partial_{t}c_{k}=g_{ij}\tilde{\beta}^{i}\tilde{\beta}^{j}\,.
\ee
Our construction gives a diagrammatic representation of it within the loop expansion.
It is also clear now what we meant by diagonal or nondiagonal contributions: they refer to the entries of this metric.
In principle one can evaluate all these diagrams for a generic cutoff $R_k(z)$ but this turns out to be a difficult analytical task.
In the next section we will be able to evaluate analytically one particular class of diagonal entries\footnote{One can see that in the limit $m^2\to0$ and for coupling--independent entries this are the only non--zero diagonal contributions.}.

\subsubsection{Diagonal contributions}

At $L$--loop order, 
the simplest coupling--independent diagonal contribution comes from the following diagram:  
\begin{center}
\includegraphics[scale=.5]{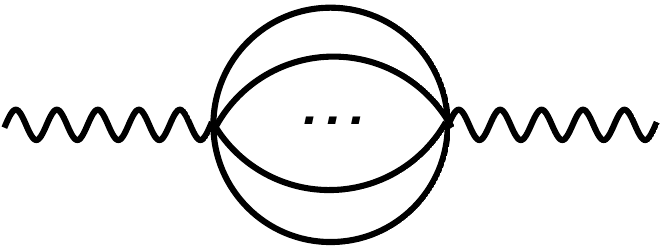}
\end{center}
corresponding to the expression:
\be
\partial_{t}\Gamma_{L,k} = -\frac{1}{2(L+1)!}\, \tilde{\beta}_{L+1}^{2}\,k^{4}\int d^{2}x\int d^{2}y\,\tau_x \tau_y \, \tilde{\partial}_{t}\left[G_{k}\left(x-y\right)\right]^{L+1}
\ee
(which generalizes equation (\ref{loop3}) of the Appendix).
In the above equation the $2(L+1)!$ comes from the symmetry factor of the diagram, and the minus sign from the fact that we are acting with an overall $\tilde{\partial}_t$.
To recover the contribution to $\partial_{t}c_k$ is simple: expand $\tau_y$ around $x$ as in equation (\ref{tauexp}), and isolate the proper term according to equation (\ref{dtcexact}).
%

To see more explicitly the form that the metric of Zamolodchikov takes, we need some preliminary results.
Using a mass cutoff $R_{k}=k^{2}$, the zero mass running renormalized propagator (\ref{prop}) will be the same
as the standard massive one, only with $k^2$ in place of the mass $m^{2}$,
and the cutoff vertices play no role.
In real space the propagator reads:
\be
G_{k}\left(x-y\right)=\frac{1}{2\pi}K_{0}\left(\left|x-y\right|\sqrt{ak^{2}}\right)\,,
\ee
where $K_{0}$ is the Bessel $K$--function of order zero. 
We introduced the parameter $a$, eventually to be sent to 1, since in this way
we have the simpler formula
\be
\tilde{\partial}_{t}f\left[R_{k}\right]=2\partial_{a}\left.f\left[ak^{2}\right]\right|_{a\rightarrow1}\,.
\ee
The different contributions are then calculated after expanding $\tau_y$
around $x$ using (\ref{tauexp}). We find:
\be
\partial_{t}\Gamma_{L,k}=\frac{k^4}{(L+1)!}\, \tilde{\beta}_{L+1}^{2} \int d^{2}x \, \tau_x \Delta \tau_x \int d^{2}y\frac{y^{2}}{2(2\pi)^{L+1}}\partial_{a}\left.\left[K_{0}\left(\left|y\right|\sqrt{ak^{2}}\right)\right]^{L+1}\right|_{a\rightarrow1}\,.
\ee
%
These diagonal terms can be written
to all orders, they give a contribution to the flow equation for $c_{k}$
of the form:
\be
\partial_{t}c_{L,k}={\cal A}_{L} \, \tilde{\beta}_{L+1}^{2}\,,
\ee
in which we defined the quantity
\be
{\cal A}_{L}\equiv\frac{3}{2^{L}\pi^{L-1}L!}\int_{0}^{\infty}dx\, x^{4}\left[K_{0}\left(x\right)\right]^{L}K_{1}\left(x\right)\,.
\ee
Note the interesting thing that contributions at loop order $L$ are proportional to the square of the beta function of the coupling $\tilde{\lambda}_{L+1,k}$.
Thus the flow of $c_k$ receives contributions from all loops (as it is inherently non-perturbative) but a given interaction starts to contribute only at a given loop order.  
All the ${\cal A}_{L}$ can be evaluated numerically and they turn out to be positive.
The numerical values of the first ${\cal A}_{L}$ are shown in table 1. Note the fast decrease relative to the one--loop value.

We can now write down the contribution of this class of diagrams to the running of the $c$--function at all loops in the $\mathbb{Z}_{2}$--symmetric case:
\be
\partial_{t}c_{k}^{\left(diagonal\right)}=\sum_{i=1}^{\infty}{\cal A}_{2i-1} \, \tilde{\beta}_{2i}^{2}\,,
\ee
which also gives the explicit form for the diagonal entries of the  Zamolodchikov metric.
Since this sum is manifestly positive, we can say that the $c$--theorem is satisfied to all loops by the diagonal terms considered.

\begin{table}
\begin{centering}
\begin{tabular}{|c|c|c|c|c|c|c|}
\hline 
{\small ${\cal A}_{1}$} & {\small ${\cal A}_{2}$} & {\small ${\cal A}_{3}$} & {\small ${\cal A}_{4}$} & {\small ${\cal A}_{5}$} & {\small ${\cal A}_{6}$} & {\small ${\cal A}_{7}$}\tabularnewline
\hline 
{\small 1} & {\small 0.0182} & {\small $4.778 \times 10^{-4}$} & {\small $1.485 \times 10^{-5}$} & {\small $5.066 \times 10^{-7}$} & {\small $1.825 \times 10^{-8}$} & {\small $6.8 \times 10^{-10}$}\tabularnewline
\hline 
\end{tabular}
\par\end{centering}
\caption{First few numerical values of ${\cal A}_{L}$.}
\end{table}
%

%
%

As we have seen previously,  the entries of Zamolodchikov's metric contain a coupling--independent piece, plus further pieces proportional to increasing powers of the coupling constants, as we increase the loop order.
The positivity properties of the metric are far from trivial when all these terms are involved.
However, when the couplings are sufficiently small, the positivity will be determined solely by the coupling independent terms. 


\subsubsection{Non--unitary theories}

Finally we make a comment on when the $c$--theorem is not satisfied, i.e. the case when $\partial_t c_k <0$.
We know that the $c$--theorem does not hold without the unitarity assumption \cite{Zamolodchikov:1986gt}.
This can indeed be checked explicitly.
It's easy to see that when one considers interactions with complex couplings then the coefficients in the loop expansion turn negative.
For instance, one notable example is the Lee-Yang model \cite{mussardo}, in which one introduces the
non--unitary complex interaction:
\be
S_{LY}[\phi]=\int d^2x \left[ \frac{1}{2}\phi \Delta \phi + ig \phi^3 \right] \,.
\ee
A simple analysis reveals that this interaction contributes to the running
of $c_{k}$ through the following diagram:
\begin{center}
\includegraphics[scale=.65]{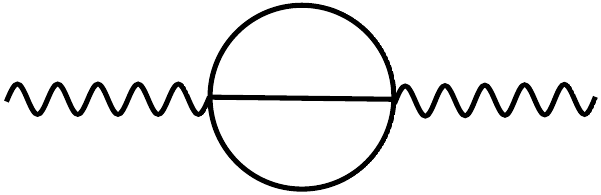}
\end{center}
which turns out to have the wrong sign to be consistent with the $c$--theorem:
%
%
\be
\partial_t c_k = - \mathcal{A}_2\, \tilde{\beta}_3^2<0\,,
\ee
since $ \mathcal{A}_2 > 0$, as reported in table 1.

\section{The $c$--function and Newton's constant}

In this section we derive an interesting relation between the $c$--function and the matter induced beta function of Newton's constant.\footnote{In what follows we identify the Newton's constant as the coupling in front of the Ricci scalar. In a non linear sigma model on curved target space this coupling is equivalently identified as the dilaton constant mode.}
This can then be used to obtain another form of the flow of the central charge $\partial_t c_k$.

\subsection{Relation between $c_k$ and $\beta_{G_k}$}
To obtain this relation we need to consider what happens when in equaiton (\ref{nonlocEAA0}) we set $\mathcal{O}= R\,$.
Since the coupling constant of the invariant $\int \sqrt{g} R$ is $-\frac{1}{16\pi G_k}$, where $G_k$ is the running Newton's constant,
one finds, for the gravitational part of the EAA, the following form\footnote{We need here $1/4$ instead of $1/2$ because of the the further symmetry we have in exchanging the two $R$s}:
\be
\Gamma_{k}[0,g] =  \int d^2x\sqrt{g}\left[-\frac{1}{16\pi G_{k}}R-\frac{1}{4}\partial_{t}\left(-\frac{1}{16\pi G_{k}}\right)R\frac{1}{\Delta}R+...\right]
 \,.
\ee
We recognize that the Polyakov term above is the same that we included in our general anstaz for the EAA (\ref{nonlocEAA}). 
Thus we infer that there is a relation between the beta function of Newton's constant and the running $c$--function:
\be
\partial_{t}\left(-\frac{1}{16\pi G_{k}}\right)=\frac{\mathcal{C}_k}{24\pi}\,.
\label{cG}
\ee
This is a nontrivial statement by itself. It tells us that the running
$c$--function for a certain matter field type can also be computed
from the contributions of that kind of matter to the beta function
of Newton's constant. In fact a derivative of (\ref{cG}) with respect to the RG scale gives (remember that $\partial_t \mathcal{C}_k=\partial_t c_k$):
\be
\partial_t c_k = \frac{3}{2 G_k^2}\left( \partial_t \beta_{G_k}-2\frac{\beta_{G_k}^2}{G_k}\right)\,,
\ee
where $\beta_{G_k}\equiv\partial_t G_k$ is the Newton's constant beta function. 
We will check the consistency of relation (\ref{cG}) in the case of a minimally coupled and a self--interacting scalar.

\subsection{Minimally coupled scalar}
\begin{figure}
\begin{centering}
\includegraphics[scale=0.3]{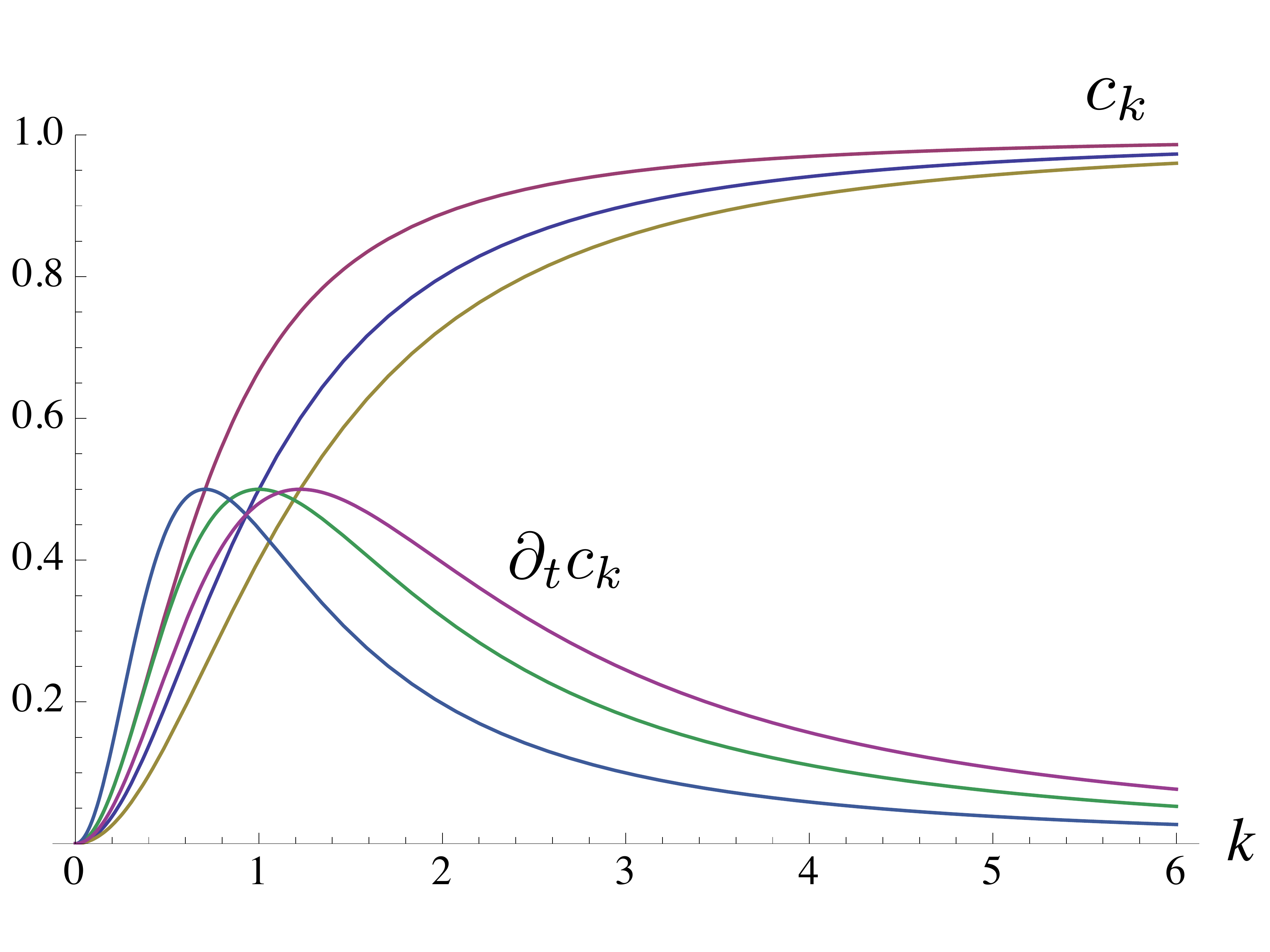} 
\caption{$c_k$ and $\partial_t c_k$ as a function of $k$ for a massive deformation of a minimally coupled scalar. Mass ($a=1$), optimized ($a=1$) and exponential ($a=1$, $b=1$) cutoffs (upper curves),  exponential ($a=1$, $b=\frac{1}{2}$) cutoff (middle curves), exponential ($a=1$, $b=\frac{3}{2}$) cutoff (lower curves). In all cases we set $m^2=1$. \label{miniscalar}}
\end{centering}
\end{figure}

Consider a minimally coupled scalar describing a massive deformation of the Gaussian fixed point as discussed in section 5.1.1.
The action is given in (\ref{scalar}) and the exact flow equation (\ref{exactflow}) for this case reads:
\be
\partial_t \Gamma_k[\varphi,g]=\frac{1}{2}\textrm{Tr}\frac{\partial_t R_k(\Delta)}{\Delta+m^2+R_k(\Delta)}\,.
\label{exactflow_scalar}
\ee
Note that the dilaton plays no role now, since we are free to set $\tau=0$.
Instead, to find $c_k$ using (\ref{cG}) we need to extract the terms in the trace on the r.h.s. of (\ref{exactflow_scalar}) that are proportional to the invariant $\int \sqrt{g} R$.
As usual, this can be done using the heat kernel expansion \cite{Codello:2008vh}.
Defining $h_k(z)=\frac{\partial_t R_k(z)}{z+m^2+R_k(z)}$, one finds:
\be
\frac{1}{2}\textrm{Tr}\,h_k(\Delta) \Big|_{ \int \sqrt{g} R } = \frac{1}{8\pi} \frac{1}{6} h_k(0)\int d^2x \sqrt{g} R\,,
\ee
which, when compared with the scale derivative of $-\frac{1}{16\pi G_{k}}\int \sqrt{g} R$ on the l.h.s. of (\ref{exactflow_scalar}), gives:
\begin{equation}
\partial_{t}\left(-\frac{1}{16\pi G_{k}}\right) = \frac{1}{8\pi}\frac{1}{6} h_k(0)\,.
\end{equation}
Thus our formula (\ref{cG}) leads to\footnote{We are not using Weyl quantization so $\mathcal{C}_k=c_k$}:
\begin{equation}
c_{k} = \frac{1}{2} h_k(0)\,.
\label{c_scalare}
\end{equation}
Note that this relation is valid for arbitrary cutoff function $R_k(z)$, as opposed to the result of section 5.1.1 valid only for the mass cutoff.
For both the mass cutoff $R_k(z)=ak^2$ and the optimized cutoff $R_k(z)=a(k^2-z)\theta (k^2-z)$ we find the following form:
\be
c_{k}=\frac{a k^2}{a k^2+m^2}\,.
\ee
For the exponential cutoff $R_k(z)=\frac{a z}{e^{b z/k^2}-1}$, with parameters $a$ and $b$, we find:
\be
c_{k}=\frac{a k^2}{a k^2+b m^2}\,.
\label{c_miniscalar}
\ee
In all cases and for all values of the parameters $a$ and $b$ we find that $c_{UV}=1$ and $c_{IR}=0$ as expected.
A derivative of (\ref{c_miniscalar}) gives the flow of the $c$--function:
\be
\partial_t c_k =\frac{2 a b k^2 m^2}{\left(a k^2+b m^2\right)^2}\,.
\ee
The interpolating $c_k$ of equation (\ref{c_miniscalar}) and the flow of the last equation are shown in figure \ref{miniscalar}.
We clearly see that the flow is scheme dependent, but the integral of it along a trajectory, giving $\Delta c$, is universal.

\subsection{Self--interacting scalar}
\begin{figure}
\begin{centering}
\includegraphics[scale=0.6]{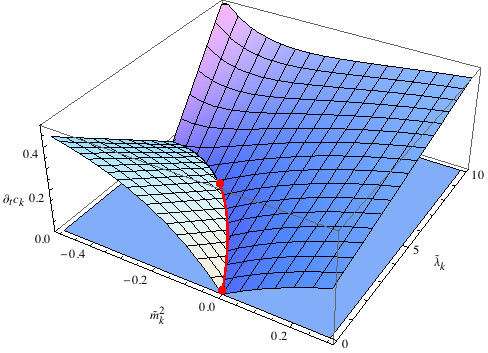} 
\caption{$\partial_t c_k$ in the $(\tilde{m}_k^2,\lambda_k)$ plane according to (\ref{c_LPA_cG_1}) for $a=1$ and $b=\frac{1}{2}$ . We marked with a red dot the position of the Gaussian and Ising fixed points and the trajectory--$III$ connecting them.
One can note that the trajectory connecting the two fixed points lies along a "valley" where the variation in the flow of the central charge is zero.\label{fig: plot_c_flow_cG}}
\end{centering}
\end{figure}
We consider now an interacting scalar, i.e. the LPA action (\ref{LPAansatz}) of section 5.2. 
We can obtain $c_k$ directly from equation (\ref{c_miniscalar}) by just making the replacement $m^2 \to V''_k(\varphi_0)$:
\be
c_{k}=\frac{a k^2}{a k^2+b V''_k(\varphi_0)}\,.
\ee
A scale derivative now gives:
\be
\partial_t c_k = -\frac{a b k^2 \left( \partial_t V''_k(\varphi_0)-2V''_k(\varphi_0) \right)}{\left(a k^2+b V''_k(\varphi_0) \right)^2}\,.
\label{c_LPA_cG_0}
\ee
We need to decide the value of $\varphi_0$ where to evaluate this expression.
In this case it is important to distinguish the ordered from the broken phase.
If the running effective potential has the polynomial form (\ref{LPA_Taylor}), then we have $\varphi_0=0$ in the ordered phase and $\varphi_0=\pm \sqrt{6 m_k^2 / \lambda_k}$ in the broken phase, the two phases being separated by trajectory--$III$ and its continuation.
Inserting these expressions in (\ref{c_LPA_cG_0}) gives the following form for the flow of $c_k$:
\be
\partial_t c_k =
\left\{ \begin{array}{cc}
-\frac{ab\,\partial_{t}\tilde{m}_{k}^{2}}{(a+b\,\tilde{m}_{k}^{2})^{2}} & \textrm{ordered phase}\\
\\
\frac{2ab\,\partial_{t}\tilde{m}_{k}^{2}}{(a-2b\,\tilde{m}_{k}^{2})^{2}} & \textrm{broken phase}\,.
\end{array}\right.
\label{c_LPA_cG_1}
\ee
As shown in figure \ref{fig: plot_c_flow_cG}, the flow (\ref{c_LPA_cG_1}), even if not proportional to the square of the dimensionless beta function, is positive $\partial_t c_k \geq 0$ in the $(\tilde{m}_k^2,\lambda_k)$ plane:

This calculation represents a non--trivial check of relation (\ref{cG}) and shows how this relation can be used explicitly to compute $c_k$ in a given truncation by means of heat kernel techniques.

\section{Conclusions}

In this work we have explored a new way to study the flow of the $c$--function
within the framework of the functional RG based on the effective average action (EAA).
This function interpolates between the UV and IR central charges of the corresponding
CFTs and is thus a global feature of the flow,
related to the integration of it along a trajectory connecting two fixed points, independent of scheme ambiguities.

%
Our main result is an RG exact equation for the running $c$--function based on the identification
of it with the coefficient of the running Polyakov action.
This equation relates the flow of the central charge to the exact flow of the EAA.
%
%
To solve the equation for non--trivial cases we built a suitable ansatz requiring the
EAA to reproduce generic features of QFTs, namely the scale and the conformal anomalies.
In its own right this is an interesting result since it teaches us 
that a consistent ansatz for the EAA off criticality should include 
some nonlocal terms proportional to beta functions.
%
%
%
Of course we do not claim full generality for this ansatz, but we found that it is sufficiently
accurate to trigger the flow of  the $c$--function in non--trivial cases.
Explicit computations, within the local potential approximation and the loop expansion, have been presented in section 5 showing the compatibility of our framework with the $c$--theorem.

Moreover we have put forward a relation between the beta function of Newton's constant and
the running conformal anomaly. This relation comes from internal consistency of the generic ansatz 
for the EAA we proposed and allows us to use heat kernel techniques to compute the RG running of the $c$--function.
We also checked this other relation in explicit cases, finding it consistent.
Nevertheless we point out that our analysis is not complete.
The works \cite{Osborn:1991gm,Friedan:2009ik} highlight that there are some subtleties related to the definition of the  $c$--function. A complete mapping between the local RG approach and the fRG is still lacking and further study is needed in this direction.
Another issue, which has not been touched at all, is the generalization of these ideas to the higher dimensional case, in particular $d=4$ where one can consider similar constructions for the $a$--function \cite{Osborn:1991gm,Fortin:2012hn,Luty:2012ww}, which we leave to future work.
%
%
%

\pagebreak

\appendix
\section{Loop expansion from the fRG}

%
%
The exact flow equation (\ref{exactflow}) satisfied by the EAA can be solved iteratively.
If we choose as seed for the iteration the bare action,
then the iteration procedure reproduces the renormalized loop expansion \cite{Litim:2001ky,Codello:2013bra}.

One starts with $\Gamma_{0,k} \equiv S_\Lambda$, where
$S_{\Lambda}$ is the UV or bare action, and sets up an iterative solution
(the subscript $0$ indicates the order of the iteration, $\Lambda$ is the UV cutoff and $k$
is the RG scale) by plugging $\Gamma_{0,k}$ into the r.h.s.
of the flow equation and integrates the resulting differential equation
with the boundary condition $\Gamma_{1,\Lambda}=S_{\Lambda}$.
The solution $\Gamma_{k,1}$
is then plugged back into the r.h.s. of the flow equation and the procedure
is be repeated.

%
To see this let us introduce $\hslash$ as a loop counting parameter and expand the EAA:
\be
\Gamma_{k}=S_{\Lambda}+\sum_{L=1}^{\infty}\hslash^{L}\Gamma_{L,k}\,.
\ee
The bare action is $k$--independent $\partial_{t}S_{\Lambda}=0$.
The exact flow equation (\ref{exactflow}) now takes the form:
\be
\hslash\,\partial_{t}\Gamma_{1,k}\left[\varphi\right]+\hslash^{2}\partial_{t}\Gamma_{2,k}\left[\varphi\right]+...=\frac{\hslash}{2}\mbox{Tr}\frac{\partial_{t}R_{k}}{S_{\Lambda}^{\left(2\right)}\left[\varphi\right]+R_{k}+\hslash\,\Gamma_{1,k}^{\left(2\right)}\left[\varphi\right]+
\hslash^{2}\Gamma_{2,k}^{\left(2\right)}\left[\varphi\right]+...}\,.
\label{looperge}
\ee
The original flow equation (\ref{exactflow}) is finite both in the UV and IR: to maintain
these properties the bare action $S_{\Lambda}$ has to contain counterterms to cancel
the divergencies that may appear in the $\Gamma_{L,k}$. Thus we define:
\be
S_{\Lambda}=S_{0}+\sum_{L=1}^{\infty}\hslash^{L}\Delta S_{L,\Lambda}\,,
\label{counterterms}
\ee
where each counterterm $\Delta S_{L,\Lambda}$ is chosen to cancel the divergent part of $\Gamma_{L,0}$.
Since this divergent part is the same as the divergent part of $\Gamma_{L,k}$ (we refer to \cite{Codello:2013bra} for more details on this point),
this choice renders the denominator of (\ref{looperge}) finite.
Here $S_0$ is the renormalized action, i.e. the bare action with renormalized fields, masses and couplings.
From (\ref{looperge}) we can read off the flow of the $L$--th loop contribution:
\be
\partial_{t}\Gamma_{L,k}\left[\varphi\right]=\frac{1}{\left(L-1\right)!}\frac{\partial^{L-1}}{\partial \hslash^{L-1}}\left.\frac{\partial_{t}\Gamma_{k}\left[\varphi\right]}{\hslash}\right|_{\hslash\rightarrow0}\,.
\label{loopgen}
\ee
The one--loop equation is straightforward:
\be
\partial_{t}\Gamma_{1,k}\left[\varphi\right]=\frac{1}{2}\mbox{Tr}\, G_{k}\left[\varphi\right]\partial_{t}R_{k}\,,
\label{loop1}
\ee
where the $k$--dependent renormalized propagator,
\be
G_{k} [\varphi ]=\frac{1}{S_{0}^{\left(2\right)}\left[\varphi\right]+R_{k}}\,,
\label{prop}
\ee
depends on $k$ only trough the cutoff $R_k$. Thus, within the loop expansion,
the operator $\tilde{\partial}_t$, introduced in section 3.2, is equivalent to $\partial_t$.
%

We can integrate the one--loop flow equation (\ref{loop1}) between the UV and IR scales.
We choose the UV initial condition $\Gamma_{L,\Lambda}=0$ for $L>0$ since the UV action is just the bare action.
We find:  
\begin{eqnarray}
\Gamma_{1,k} & = & -\int_{k}^{\Lambda}\frac{dk^{\prime}}{k^{\prime}}\partial_{t^{\prime}}\Gamma_{1,k^{\prime}}=-\frac{1}{2}\int_{k}^{\Lambda}\frac{dk^{\prime}}{k^{\prime}}\mbox{Tr}\,G_{k^{\prime}}\partial_{t^{\prime}}R_{k^{\prime}}\nonumber\\
& = & \frac{1}{2}\int_{k}^{\Lambda}dk^{\prime}\mbox{Tr}\,\partial_{k^{\prime}}\log G_{k^{\prime}}=\left.\frac{1}{2}\mbox{Tr}\log G_{k}\right|^{\Lambda}_{k}.
\end{eqnarray}
Note that in the second line we have exchanged the order of the trace and the derivative.
This has been possible since we inserted an additional UV regulator
$\Lambda$ (one can also use dimensional regularization \cite{Codello:2013bra}).
%
In the following all manipulations are intended
with an implicit UV cutoff $\Lambda$.

We now choose $\Delta S_{L,\Lambda}=-[\Gamma_{L,0}]_{div}$ and define the renormalized one--loop contribution:
%
\be
[\Gamma_{1,0}]_{ren}\equiv \lim_{\Lambda \to \infty} \left( \Gamma_{1,k}+\Delta S_{1,\Lambda} \right) = \frac{1}{2}[\mbox{Tr}\log G_{k}]_{ren}\,.
\ee
Obviously, this limit is finite only if the theory is perturbatively renormalizable.


Now let us consider the two-loop contribution:
\be
\partial_{t}\Gamma_{2,k}=\frac{\partial}{\partial \hslash}\frac{\partial_{t}\Gamma_{k}}{\hslash}\Big|_{\hslash \to 0}=-\frac{1}{2}\mbox{Tr}\,G_{k}[\Gamma_{1,k}^{(2)}]_{ren}G_{k}\partial_{t}R_{k}=\frac{1}{2}\mbox{Tr}[\Gamma_{1,k}^{(2)}]_{ren}\partial_{t}G_{k}.
\ee
%
We can plug in the one--loop result previously found. To do that
we need to compute the Hessian $\Gamma_{1,k}^{(2)}$:
\be
\Gamma_{1,k}^{(2)}  =
 -\frac{1}{2}G_{k}S_{0}^{(3)}G_{k}S_{0}^{(3)}+\frac{1}{2}S_{0}^{(4)}G_{k}\,,
\ee
where we suppressed all indices. Using the above equation we get:
\begin{eqnarray}
\partial_{t}\Gamma_{2,k} & = &
\frac{1}{2}\left[-\frac{1}{2}G_{k}S_{0}^{(3)}G_{k}S_{0}^{(3)}+\frac{1}{2}S_{0}^{(4)}G_{k}\right]^{ab}_{ren}\left[\partial_{t}G_{k}\right]^{ba}\nonumber\\
& = & \frac{1}{2} \partial_{t}\left[-\frac{1}{3\cdot2}G_{k}^{cd}S_{0}^{(3)ade}G_{k}^{ef}S_{0}^{(3)bfc}G_{k}^{ab}+\frac{1}{2\cdot2}S_{0}^{(4)abcd}G_{k}^{cd}G_{k}^{ab}\right]_{ren}\,,
\label{loop2}
\end{eqnarray}
where we used relations (\ref{relations}) to extract the overall scale derivative.
Integrating and renormalizing (\ref{loop1}) as before gives:
\begin{equation}
\Gamma_{2,k} = \left[-\frac{1}{12}G_{k}^{cd}S_{0}^{(3)ade}G_{k}^{ef}S_{0}^{(3)bfc}G_{k}^{ab}+\frac{1}{8}S_{0}^{(4)abcd}G_k^{cd}G_{k}^{ab}\right]_{ren}.
\end{equation}
In the limit $k \to 0$ we recovered the usual two--loop result with the correct coefficients and in (nested) renormalized form.
We can represent diagrammatically these contributions by adopting the same rules of section 3.2 with the difference that  a continuous line represents a renormalized regularized propagator and vertices are constructed from the renormalized action $S_0^{(m)}$. To each loop we associate an integration $\int d^2x$ in coordinate space or $\int \frac{d^2q}{(2\pi)^2}$ in momentum space and we act overall with  $\partial_t$. 
Proceeding along these lines all the standard loop expansion can be recovered at any loop order.
From now on, for notational simplicity we will omit to explicitly report renormalized quantities with bracket, since these can be understood from the context. 

Starting at three--loop order there are many different contributions.
Here we show how to compute the following diagram,
\begin{center}
\includegraphics[scale=.6]{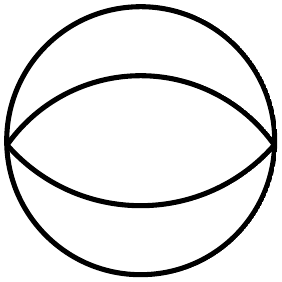}
\end{center}
that we will use and generalize in section 5.3.
We start from the following three--loop term flow:
\be
\partial_{t}\Gamma_{3,k}=\frac{1}{2}\left(G_{k}^{ab}\Gamma_{1,k}^{(2)bc}G_{k}^{cd}\Gamma_{1,k}^{(2)de}G_{k}^{eg}-G_{k}^{ab}\Gamma_{2,k}^{(2)bc}G_{k}^{cg}\right)\partial_{t}R_{k}^{ga}.
\ee
We need the Hessian of the two-loop renormalized contribution,
%
%
considering that we are interested only in the three-loop contribution in which
there are two vertices $S^{(4)}_0$. Therefore we select:
\begin{eqnarray*}
\Gamma_{2,k}^{(2)mn} = \frac{1}{12}\Biggl[G_{k}^{cd}S_{0}^{(4)adem}G_{k}^{ef}S_{0}^{(4)bfcn}G_{k}^{ab}+G_{k}^{cd}S_{0}^{(4)aden}G_{k}^{ef}S_{0}^{(4)bfcm}G_{k}^{ab}\Biggr]\nonumber
\end{eqnarray*}
\be
 -\frac{1}{8}\Biggl[\left(-G_{k}^{aa_{1}}S_{0}^{(4)a_{1}a_{2}mn}G_{k,a_{2}b}\right)S_{0}^{(4)abcd}G_{k}^{cd}+G_{k}^{ab}S_{0}^{(4)abcd}\left(-G_{k}^{ca_{1}}S_{0}^{(4)a_{1}a_{2}mn}G_{k}^{a_{2}d}\right)\Biggr]\,.
\ee
So we find:
%
\begin{eqnarray*}
\partial_{t}\Gamma_{3,k} &=& \frac{1}{2}\left[G_{k}^{ab}\left(-\frac{1}{2}S_{\Lambda}^{(4)bca_{1}a_{2}}G_{k}^{a_{1}a_{2}}\right)G_{k}^{cd}\left(-\frac{1}{2}S_{\Lambda}^{(4)dea_{3}a_{4}}G_{k}^{a_{3}a_{4}}\right)G_{k}^{eg}\right.\\
&&\left.-G_{k}^{ab}\Gamma_{2,k}^{(2)bc}G_{k}^{cg}\right]\partial_{t}R_{k}^{ga}\,;
\end{eqnarray*}
recalling $-\partial_{t}G_{k}G_{k}^{(-1)}=G_{k}\partial_{t}R_{k}$
we pick up the contribution of the diagram we are interested in:
\begin{eqnarray}
\partial_{t}\Gamma_{3,k} & = & -\frac{1}{2}G_{k}^{qm}\left(\frac{1}{6}G_{k}^{cd}S_{0}^{(4)adem}G_{k}^{ef}S_{0}^{(4)bfcn}G_{k}^{ab}\right)(-\partial_{t}G_{k}^{nr})G_{k}^{(-1)rq}+\cdots\nonumber\\
 & = & -\frac{1}{2}\left(\frac{1}{6}G_{k}^{cd}S_{0}^{(4)adeq}G_{k}^{ef}S_{0}^{(4)bfcn}G_{k}^{ab}\right)(-\partial_{t}G_{k}^{nq})+\cdots\nonumber\\
 & = & \partial_{t}\left[\left(-\frac{1}{4\cdot12}\right)G_{k}^{qm}G_{k}^{cd}S_{0}^{(4)adem}G_{k}^{ef}S_{0}^{(4)bfcn}G_{k}^{ab}G_{k}^{nq}\right]+\cdots\,,
 \label{loop3}
\end{eqnarray}
where we used the cyclicity of the trace. Note that the symmetry factor
of the three--loop contribution to the effective action is automatically recovered.
Similarly one can easily obtain all the higher loop diagrams of this form.
%

\pagebreak

\end{document}